\documentclass[a4paper,twocolumn,accepted=2020-03-12]{quantumarticle}
\pdfoutput=1

\usepackage{graphicx}
\usepackage[T1]{fontenc}
\usepackage{amsmath, amsthm, amssymb, mathtools, dsfont}
\usepackage{graphicx}
\usepackage{dcolumn}
\usepackage{bm,bbm}
\usepackage[normalem]{ulem}
\usepackage{color}
\usepackage[usenames,dvipsnames]{xcolor}
\usepackage{esint}
\usepackage{paralist}

\usepackage{geometry}
\usepackage[utf8]{inputenc}
\usepackage[english]{babel}
\usepackage[backend=bibtex,
style=numeric,
sorting=none,
doi=true]{biblatex}
\usepackage{csquotes}
\usepackage[T1]{fontenc}
\usepackage{amsmath}
\usepackage{hyperref}

\DeclareGraphicsExtensions{.png,.pdf,.eps,.svg}
\graphicspath{ {./figures/} }
\usepackage{epstopdf}

\newcommand{\ket}[1]{| #1 \rangle}

\newcommand{\ketbra}[2]{| #1 \rangle \langle #2 |}

\newcommand{\M}{\mathbf{M}} %POVM M
\newcommand{\N}{\mathbf{N}} %POVM N
\newcommand{\iden}{\mathbbm{1}}
\renewcommand{\P}{\mathbf{P}} %proj POVM Notation
\renewcommand{\vec}[1]{\mathbf{#1}}

\newcommand{\rbracket}[1]{\left(#1\right)} %round bracket
\newcommand{\sbracket}[1]{\left[#1\right]} %square bracket
\newcommand{\cbracket}[1]{\left\{#1\right\}} %curly bracket
\newcommand{\Tr}[1]{\mbox{Tr}\left(#1\right)}
\newcommand{\eq}[1]{Eq.~\eqref{#1}}
\usepackage{graphicx}
\DeclareGraphicsExtensions{.png,.pdf,.eps}
\graphicspath{ {./figures/} }

\newcommand\numberthis{\addtocounter{equation}{1}\tag{\theequation}}

\theoremstyle{plain}

\theoremstyle{plain}

\theoremstyle{plain}

\theoremstyle{plain}

\theoremstyle{plain}

\theoremstyle{plain}

\theoremstyle{plain}

\theoremstyle{remark}
\newtheorem*{rem*}{Remark}
\newtheorem{rem}{Remark}
\usepackage{booktabs} %do pieknych tabelek
\usepackage{multirow}%do pieknych tabelek
\usepackage{subfigure}
\begin{document}
	\title{Mitigation of readout noise in near-term quantum devices \qquad by classical post-processing based on detector tomography}% Force line breaks with \\
	\author{Filip B. Maciejewski}
	\email{filip.b.maciejewski@gmail.com}
	\affiliation{ University of Warsaw, Faculty of Physics, Ludwika Pasteura 5, 02-093 Warszawa, Poland}
	\affiliation{ 
	International Centre for Theory of Quantum Technologies, University of Gdansk, Wita Stwosza 63, 80-308 Gdansk, Poland}	
	\affiliation{ Center for Theoretical Physics, Polish Academy of Sciences, Al. Lotników 32/46, 02-668 Warszawa, Poland}
	\author{Zolt\'an Zimbor\'as}
	\email{zimboras@gmail.com}
	\affiliation{ Wigner Research Centre for Physics of the Hungarian Academy of Sciences,
		H-1525 Budapest, P.O.Box 49, Hungary}
	\affiliation{ BME-MTA Lend\"ulet Quantum Information Theory Research Group, Budapest, Hungary}
	\affiliation{ Mathematical Institute, Budapest University of Technology and Economics,
		P.O.Box 91, H-1111, Budapest, Hungary}
	\author{Micha\l\ Oszmaniec}
	\email{michal.oszmaniec@gmail.com}
	\affiliation{ 
		International Centre for Theory of Quantum Technologies, University of Gdansk, Wita Stwosza 63, 80-308 Gdansk, Poland}	
	\affiliation{ Center for Theoretical Physics, Polish Academy of Sciences, Al. Lotników 32/46, 02-668 Warszawa, Poland}
	\maketitle
	\begin{abstract}
		We propose a simple scheme to reduce readout errors in experiments on quantum systems with finite number of measurement outcomes. 
		Our method relies on performing classical post-processing which is preceded by Quantum Detector Tomography, i.e., the reconstruction  of a Positive-Operator Valued Measure (POVM) describing the given quantum measurement device. 
		If the measurement device is affected only by an invertible classical noise, it is possible to correct the outcome statistics of future experiments performed on the same device.
		To support the practical applicability of this scheme for near-term quantum devices, we characterize measurements implemented in IBM's and Rigetti's quantum processors. We find that for these devices, based on superconducting transmon qubits, classical noise is indeed the dominant source of readout errors. Moreover, we analyze the influence of the presence of coherent errors and finite statistics on the performance of our error-mitigation procedure. 
		Applying our scheme on the IBM's 5-qubit device, we observe a significant improvement of the results of a number of single- and two-qubit tasks including Quantum State Tomography (QST), Quantum Process Tomography (QPT), the implementation of non-projective measurements, and certain quantum algorithms (Grover's search and the Bernstein-Vazirani algorithm). 
		Finally, we present results showing improvement for the implementation of certain probability distributions in the case of five qubits.
	\end{abstract}

	\section{Introduction}\label{sec:Introduction}
	In recent years, quantum technologies have been rapidly developing. 
	Scientists and engineers around the world share the hope and fascination caused by the possibility of creating devices that would allow for the manipulation of delicate quantum states with unprecedented precision \cite{Preskill2018}.
	Due to the advent of quantum cloud services (IBM \cite{qiskit_ref,ref_IBMb}, Rigetti \cite{ref_rigetti}, DWave \cite{ref_dwave}), 
	any researcher has a possibility to perform experiments on actual quantum devices.
	However, if one really hopes for utilizing such near-term devices for real-life applications such as quantum computation \cite{mike&ike}, quantum simulations \cite{Georgescu2014} or generating random numbers \cite{tamura2019}, experimental imperfections must be taken into account.
	Hence, to properly characterize noise occurring in the devices and to develop error correction and mitigation schemes that may help to fight it have become tasks of fundamental importance \cite{Li2017,Kandala2017,Kandala2019,Temme2017,Endo2018,Premakumar2018,Bonet-Monroig2018}.
	In the present work, we address this problem for the noise affecting quantum measurements.
	
	\begin{figure}[t!]
		\
		\centering
		\includegraphics[scale=0.5]{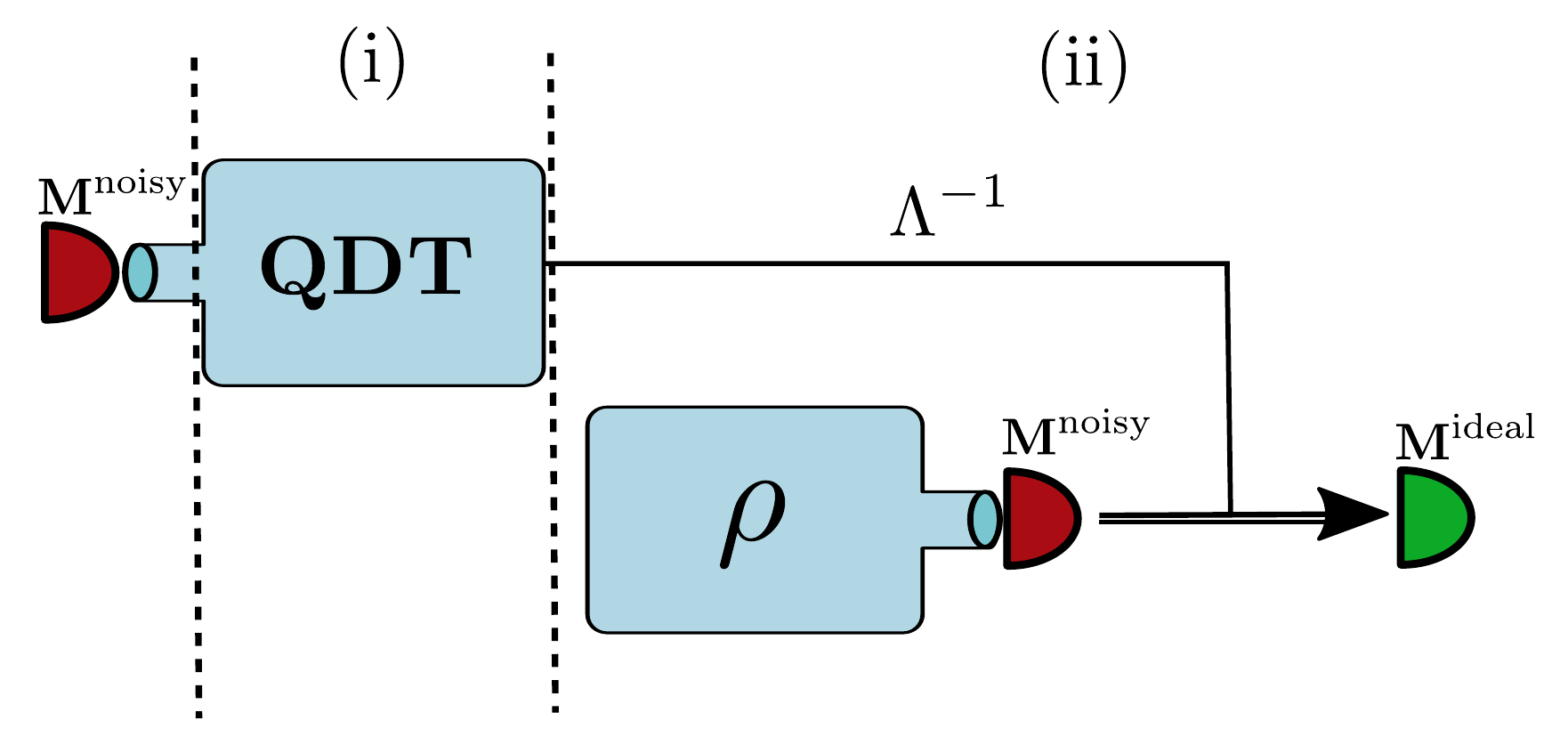}
		\caption{\label{fig:corr_scheme}
			Pictorial representation of our error-mitigation procedure. 
			(i) In the first stage, one performs the tomography of a noisy detector $\M^{\text{noisy}}$ (red semicircle). 
			(ii) In the next stage, when measuring an arbitrary quantum state $\rho$, one employs a post-processing procedure on the measured statistics through the application of $\Lambda^{-1}$, the inverse of a stochastic noise map obtained in the QDT. 
			This gives access to the statistics that would have been obtained in an ideal detector $\M^\text{\text{\text{ideal}}}$ (green semicircle).}
	\end{figure}
	In theoretical considerations about quantum information tasks, a quantum device is often assumed to perform perfect measurements. 
	In practice, this assumption may be significantly violated due to experimental imperfections and noise.
	Specifically, a broad class of errors occurring in quantum devices may be described as State Preparation And Measurement (SPAM) errors \cite{Combes2017,Sun2018}.
	It follows that if state preparation errors are negligible, one is able to reconstruct a POVM (Positive-Operator Valued Measure) associated with a given measurement device, and therefore to infer measurement errors.
	A standard method of such reconstruction is to perform Quantum Detector Tomography (QDT) \cite{Lundeen2008}. 
	Results of the QDT may be used to compare the reconstructed POVM with the ideal one \cite{Lundeen2008} and to analyze the nature of measurement errors \cite{Zhang2012a,Zhang2012b,Renema2012,Blumoff2016}. 
	In this work, we propose to do more -- we show how to use the knowledge about the POVM associated with a measurement device to correct the results obtained in further experiments performed on this device.
	Such a correction is possible provided the noise which affects the measurement is of a classical stochastic type and non-degenerate (invertible).
	In other words, such type of noise is equivalent to invertible classical post-processing of experimental statistics, and it may also be inverted solely by classical post-processing.
	The general idea of mitigating the effects of such a noise is the following.
	Assuming that one has access to the aforementioned invertible stochastic map describing a noise, it is straightforward to use the (generally non-stochastic) inversion of this map to classically reverse effects of noise simply by multiplying the vector of experimental statistics (obtained in a further experiment) by this inverted matrix, see Fig.~\ref{fig:corr_scheme}. 
	The main aim of this work is to present such a classical correction scheme together with the analysis of its accuracy.
	Specifically, we show how the deviations from classical noise model and finite-size statistics affect our procedure, by providing upper bounds for the distance of the corrected probability distribution from the ideal noise-reversed scenario.

	Besides the theoretical description, we also test our procedure experimentally. First, we present data that suggests that IBM's and Rigetti's quantum processors are affected by a significant readout noise that can be described (to a good approximation) by the classical model characterized above. 
	Both architectures consist of superconducting transmon qubits \cite{Koch2007}, which may suggest that the classical noise may be the dominant form of measurement noise in such devices.
	Second, we test the correction scheme on the IBM's five-qubit device \textit{ibmqx4} and conclude that indeed it significantly compensates for the effects of this noise for several one- and two-qubit experiments, including Quantum State Tomography (QST) \cite{mike&ike}, Quantum Process Tomography (QPT) \cite{mike&ike}, the implementation of non-projective measurements \cite{Oszmaniec2018}, and two quantum algorithms (Grover's search \cite{Grover1996} and the Bernstein-Vazirani algorithm \cite{Bernstein1993}). 
	Furthermore, we test our scheme for the $5$-qubit experiments concerning the implementation of certain probability distributions. 
	
	We will now comment on related works. 
	There is a variety of recent research concerning the mitigation of different types of errors in contemporary quantum devices \cite{Li2017,Kandala2017,Kandala2019,Temme2017,Endo2018,Premakumar2018,Bonet-Monroig2018}.	
	In particular, in the Appendix of work \cite{Temme2017}, authors assume the uncorrelated, classical noise model for readout errors and state that they use measurement calibration data to correct 'assignment errors'. 
	We believe that this might refer to procedure-related to ours (under aforementioned assumptions).
	Similarly, in work \cite{Kandala2019} authors mention in the Appendix that they use 'readout calibration' to correct expectation values for assignment infidelity.

	Quantum Detector Tomography for superconducting qubits, along with other characterizations of measurement errors, was presented in the Ref.~\cite{Blumoff2016} for Rigetti devices. 
	Characterization of SPAM errors for IBM and Rigetti devices has been presented in Ref.~\cite{Sun2018}.
	
	The paper is organized as follows. In Section~\ref{sec:theory}, we present the necessary theoretical concepts, including the POVMs formalism, the QDT scheme and the description of classical noise. Section~\ref{sec: Main} is devoted to the main idea of this work; it consists of a detailed description of the statistics correction procedure, which is preceded by stating the necessary assumptions. 
	In Section~\ref{sec: Errors}, we analyze how violations of the assumptions affect our correction procedure.
	Section~\ref{sec::Summary} contains a summary of the correction procedure in the form of pictorial representation.
	The scheme presents practical steps that need to be done in the case of noisy projective measurement in the computational basis.
	In Section~\ref{sec: Validation}, we present experimental results from IBM's and Rigett's devices that provide insight into the magnitude of SPAM errors. 
	Section~\ref{sec: Applications} consist of experimental results from applications of our correction scheme for exemplary quantum information tasks in IBM's five-qubit device.
	Finally, we present the conclusions and some possible future research directions in Section~\ref{sec: Conclusions}.

	\section{Theoretical background}\label{sec:theory}
	In this section, the necessary theoretical background  and mathematical tools are shortly reviewed. 
	First, we define the notion of Positive-Operator Valued Measures, which are useful tools for modeling measurement noise in quantum devices.
	Second, we describe a Quantum Detector Tomography procedure, which allows reconstructing a POVM associated with a given device.
	Then we discuss a measure of the distance between quantum measurements known as the operational distance.
	Finally, we use the formalism of POVMs to precisely define classical noise affecting measurements, which will be used in  Section~\ref{sec: Main}  to derive our correction procedure.
	
	\subsection{Mathematical description of quantum measurements}
	We start by a mathematical description of a generalized quantum measurement modeled by a Positive-Operator Valued Measure \cite{Peres2006}. 
	A POVM with $n$ outcomes on a $d$-dimensional Hilbert space may be described by an $n$-element vector $\M$ of operators $M_i$ (called \textit{effects}), such that
	\begin{align}	\label{povm_def}
	\M&=\rbracket{M_1 , M_2, \dots,M_n }^{\text{T}},		
	&\forall i\ M_{i}&\geqslant\ 0 ,&\sum_{i=1}^{n}\ M_{i}=\iden,
	\end{align}
	where $M_i$'s are represented by $d\times d$ semi-definite positive matrices, $\iden$ is the operator identity and $\text{T}$ denotes transposition which we use because the column form of POVMs will be useful later.
	If a quantum system was initially described by a state $\rho$, the probability of obtaining the outcome associated with the effect $M_i$ is given by  \textit{Born's rule} 
	\begin{align}\label{eq::born_rule}
	p\rbracket{i|\rho }=\Tr{ \rho M_i }.
	\end{align}
	In other words, probabilities of obtaining particular outcomes are equal to the expectation values of the associated effects.
	In quantum information protocols, the considered measurements  are often \textit{projective} \cite{mike&ike}, which means that the effects $M_{i}$ fulfill the additional requirements $M_{i} M_{j}=\delta_{i,j}\ M_{i}$. 
	
	In the experimental part of this work, we will focus on projective measurements and their noisy versions, since this fits the set-up of the IBM Q devices. Nevertheless, all our theoretical considerations, including the correction procedure, are formulated for generalized measurements given by Eq.~\eqref{povm_def}.

	\subsection{Quantum Detector Tomography}
	Characterization of quantum devices requires knowledge of the POVM associated with a given measurement performed by the device. The procedure for obtaining this is known as Quantum Detector Tomography (QDT) \cite{Lundeen2008}. The general idea of QDT is as follows. 
	Recall that Born's rule associates the probabilities of particular outcomes with the expectation values of effects.
	If one performs multiple experiments on the set of quantum states which form a basis for Hermitian operators, then one may use the obtained statistics to decompose all effects on that basis via Born's rule.
	In the case of a $d$-dimensional Hilbert space, such a basis is $d^2$ dimensional, and this is the minimal number of different experiments (different quantum states) that must be performed to reconstruct a POVM.

	Naturally, in practice one will not have access to perfect statistics, nor to the perfectly prepared quantum states. 
	This may cause that the simple tomographic method returns nonphysical, non-positive POVM elements.
	To cope with this issue, the method of Maximum Likelihood Estimation (MLE) may be used, which asserts the positivity of the reconstructed quantum measurements \cite{Hradil2004, Fiurasek2001}. 
	In order to reconstruct POVMs from experimental data, we have thus implemented an iterative algorithm from \cite{Hradil2004} which converges to the MLE estimator.

	\subsection{Distances}
	After reconstructing the POVMs, it is useful to have a distance measure between them in order to compare the reconstructed quantum measurements with the ideal detectors and, naturally, to compare detectors with each other.
	Distances between quantum measurements are usually closely related to the probability distributions that they generate via Born's rule.
	Therefore, first, we need a measure of the distance between classical probability distributions.
	In this work, we will be using so called Total-Variation (TV) distance, which for two arbitrary probability vectors $\vec{p}=\rbracket{p_1,\ \dots, p_n}^{\text{T}}$ and $\vec{q}=\rbracket{q_1,\ \dots, q_n}^{\text{T}}$, TV distance between them is defined as
	\begin{equation}\label{eq::tv_distance}
	D_{TV}\rbracket{\vec{p},\vec{q}}\coloneqq\frac{1}{2}||\vec{p}-\vec{q}||_{1}=\frac{1}{2}\sum_{i=1}^n |p_i-q_i|\ ,
	\end{equation}
	where $||.||_{1}$ denotes $l_1$ norm. Total-variation is a very stringent measure of distance between probability distributions. It plays important role in proposals for attaining quantum computational advantage \cite{Harrow2017,Pashayan2017} and  can be also easily utilized to control expectation values of (classical) observables defined on the sample space of interest.

	TV-distance is related to the \textit{operational distance} \cite{Navascues2014,Puchala2018,Puchala2018b} between arbitrary POVMs $\M$ and $\N$ via the equality
	\begin{align}\label{eq::DopDtv}
	D_{op}(\M,\N)=\max_{\rho} D_{TV}\rbracket{\vec{p}^{\M},\vec{p}^{\N}},
	\end{align}
	where maximization goes over all quantum states $\rho$, and $\vec{p}^{\M/\N}$ denote probability distributions generated by measurement of quantum state via $\M/\N$, i.e.,
	\begin{align}\label{eq::prob_vec}
	\vec{p}^{\M}=\rbracket{\Tr{\rho M_1},\dots,\Tr{\rho M_n}}^{\text{T}}.
	\end{align}
	Hence, the operational distance is in fact the worst-case scenario of the distance between probability distributions generated by performing those measurements. 
	It can be shown \cite{Navascues2014} that operational distance may be calculated as
	\begin{align}\label{eq::op_distance}
	D_{op}(\M,\N)=\max_{x}\ ||\sum_{i\in x}\left(M_{i}-N_{i}\right)||_\infty,
	\end{align} 
	where the maximization is over all subsets of indices enumerating effects (i.e., all possible sets of outcomes) and $||.||_\infty$ denotes operator norm\footnote{The operator norm is defined in standard way, i.e., for the operator $A$ acting on the vector space $V$, the operator norm is equal to $||A||_\infty=\sup_{v} \cbracket{||Av||_\infty: \ v\in V,\ ||v||_\infty=1}$.}. 
	Operational distance between POVMs\footnote{We note that operational distance can be also be upper-bounded by another quantity used in the quantum information theory, namely the diamond norm of a difference of POVMs, which can be calculated in terms of Semidefinite Programming \cite{Watrous2018}. } has an interesting operational interpretation through the formula $D_{op}(\M,\N)=2p_{\mathrm{disc}}(\M,\N) -1$, where $p_{\mathrm{disc}}(\M,\N)$ is the optimal probability of distinguishing between measurements $\M$ and $\N$ (without using entanglement) \cite{Puchala2018}.

	%\subsection{Visualization of measurements}
	%\comfilip{ADD}

	\subsection{Classical noise affecting measurements}\label{sec::classical_noise}
	
	Now we will describe the model of classical noise that we are considering in this work. 
	Let us denote by $\M^{\text{\text{ideal}}}$ an $n$-outcome quantum measurement that in theory should be associated with our measurement device.
	In practice, due to the presence of noise, \textit{real} measurement describing our device is some POVM $\M^{\text{exp}}$.
	In this model, we assume that the relation between $\M^{\text{\text{ideal}}}$ and $\M^{\text{exp}}$ is given by a (left) stochastic, invertible map $\Lambda$, whose element $\Lambda_{i,j}\coloneqq p\rbracket{i|j} \in \sbracket{0,1}$ are defined by equation
	\begin{align}\label{eq::classical_noise}
	\forall i\  M^{\text{exp}}_{i}=\sum_j p\rbracket{i|j}M^{\text{\text{ideal}}}_j.
	\end{align}
	The  left-stochastic property of $\Lambda$ means that $\sum_i \Lambda_{i,j}= \sum_i p\rbracket{i|j}=1$.
	Note that this property asserts that $\M^{\text{exp}}$ is a proper POVM.
	The above equation is somewhat abstract since it provides a description of the noise on the measurement operators level. 
	In order to find its operational interpretation, let us use the fact that since the noise affects only the measurement process, \eq{eq::classical_noise} is fulfilled for an arbitrary quantum state which is measured on the noisy device.
	For an arbitrary quantum state $\rho$, let us denote the ideal vector of probabilities that one would've obtained in the ideal device as $\vec{p}^{\text{\text{ideal}}}$, which should be understood as in Eq~\eqref{eq::prob_vec}.
	Due to the linearity of Born's rule, from \eq{eq::classical_noise} it follows that the vector of probabilities $\vec{p}_{\text{exp}}$ obtained in an experiment on a noisy device is given by
	\begin{align}\label{eq::statistics_noise}
	\vec{p}_{\text{exp}}=\Lambda \vec{p}_{\text{\text{ideal}}}.
	\end{align}
	This allows us to give an intuitive interpretation of classical noise.
	Namely, the application of such noise is equivalent to the classical post-processing of the statistics \ that one would have obtained in an ideal scenario \cite{Haapasalo2012}. Such interpretation has been used previously in the context of the implementation of general POVMs by projective measurements (see works \cite{Oszmaniec2017,Guerini2017,Oszmaniec2018}).

	\section{Scheme of mitigation of readout errors}\label{sec: Main} 
	In this part, we lay out the idea of our procedure to mitigate readout errors.  We first state the assumptions under which the method works perfectly. Then, we formally formulate the method itself. Finally, we compute operational distances (see Eq. \eqref{eq::op_distance}) between ideal and noisy projective measurements for (i) single-qubit classical readout errors and (ii) uncorrelated classical errors affecting multi-qubit projective measurements. These numbers give us an indication of the magnitude errors that can be corrected by our error mitigation technique in realistic devices.
	
	\subsection{Assumptions}
	Our error-mitigation scheme relies on the following assumptions:
	\begin{enumerate}
		\item (\textbf{Infinite statistics}) We have access to the statistics given by Born's rule (Eq.~\eqref{eq::born_rule}).
		\item (\textbf{Classical noise}) The measurement noise occurring in the device has the form of a classical noise described in Section~\ref{sec:theory} (\eq{eq::statistics_noise}).
		\item (\textbf{Characterized detector}) We have at our disposal a perfect description of that noise.
	\end{enumerate}
	
	The approximate validity of these conditions can be motivated by the following arguments.
	Assumption of infinite statistics may be fulfilled to a good extent simply by increasing the number of experiments one gathers statistics from.
	Note that violations of this assumption introduce statistical errors that can also be taken into account.
	
	Moreover, in Section \ref{sec: Validation} (where we present results of QDT for IBM and Rigetti devices) we show that the classical noise model stated in the second condition turns out to be a dominant type of noise in the IBM quantum devices, which uses	superconducting transmon qubits.
	This suggests that classical noise may be a good measurement noise model for devices relying on similar architectures.

	Finally, the characterization of noise occurring in a detector may in practice be obtained via Quantum Detector Tomography (see Section~\ref{sec:theory}). Such reconstruction is of good quality, provided one has access to high fidelity preparations of single-qubit quantum states. 
	For example, the fidelities\footnote{State preparation errors may be obtained in the approximately readout-error-independent manner via randomized benchmarking (RB) experiments \cite{Knill2008,Gambetta2012}.
		Such experiments allow to calculate the average fidelities of quantum gates (that are used to prepare quantum states) in a way that is independent of measurement errors.} 
	of single-qubit gates in IBM quantum devices are typically high (of the order of $99.6\%$ - see Section~\ref{sec: Validation}).
	
	We conclude by noting that the violations of the assumptions affect the applicability of our error-mitigation procedure. The effects of such errors are analyzed in Section~\ref{sec: Errors}.

	\subsection{Correction of the statistics}
	Let us now describe the error-mitigation procedure. 
	In fact, all we need is to notice that since the matrix $\Lambda$ representing the noise is invertible, we may simply left-multiply the \eq{eq::statistics_noise} by its inverse $\Lambda^{-1}$, obtaining
	\begin{align}\label{statistics_correction}
	\vec{p}_{\text{\text{ideal}}}=\Lambda^{-1} \vec{p}_{\text{exp}}.
	\end{align}
	What \eq{statistics_correction} tells us, is that provided the knowledge of $\Lambda$, one can simply left-mulitply $\vec{p}_{\text{exp}}$  by its inverse, in order to obtain statistics $\vec{p}_{\text{\text{ideal}}}$ that one would have obtained in the experiments performed on the ideal, non-noisy device.
	This result indeed appears very natural if one recalls the fact that such classical noise is equivalent to the classical post-processing of ideal statistics (see Section \ref{sec:theory}).

	\subsection{Correction of statistics based on Quantum Detector Tomography in (multi-)qubit devices}
	So far, we have not assumed any particular structure of the measurement $\M^{\text{ideal}}$. 
	In this subsection (and in fact throughout the most part of the paper), we focus on projective measurements in the computational basis in single-qubit and multi-qubit systems. 
	Specifically, we describe in detail how our mitigation method works in these systems. Moreover, we analytically quantify the magnitude of statistical errors that our scheme is capable of correcting under the assumption of independent classical errors affecting the measurements.
	
	\subsubsection{Single qubit}\label{sec::single_qubit_corr}
\begin{figure}
		\centering
		\includegraphics[scale=0.59]{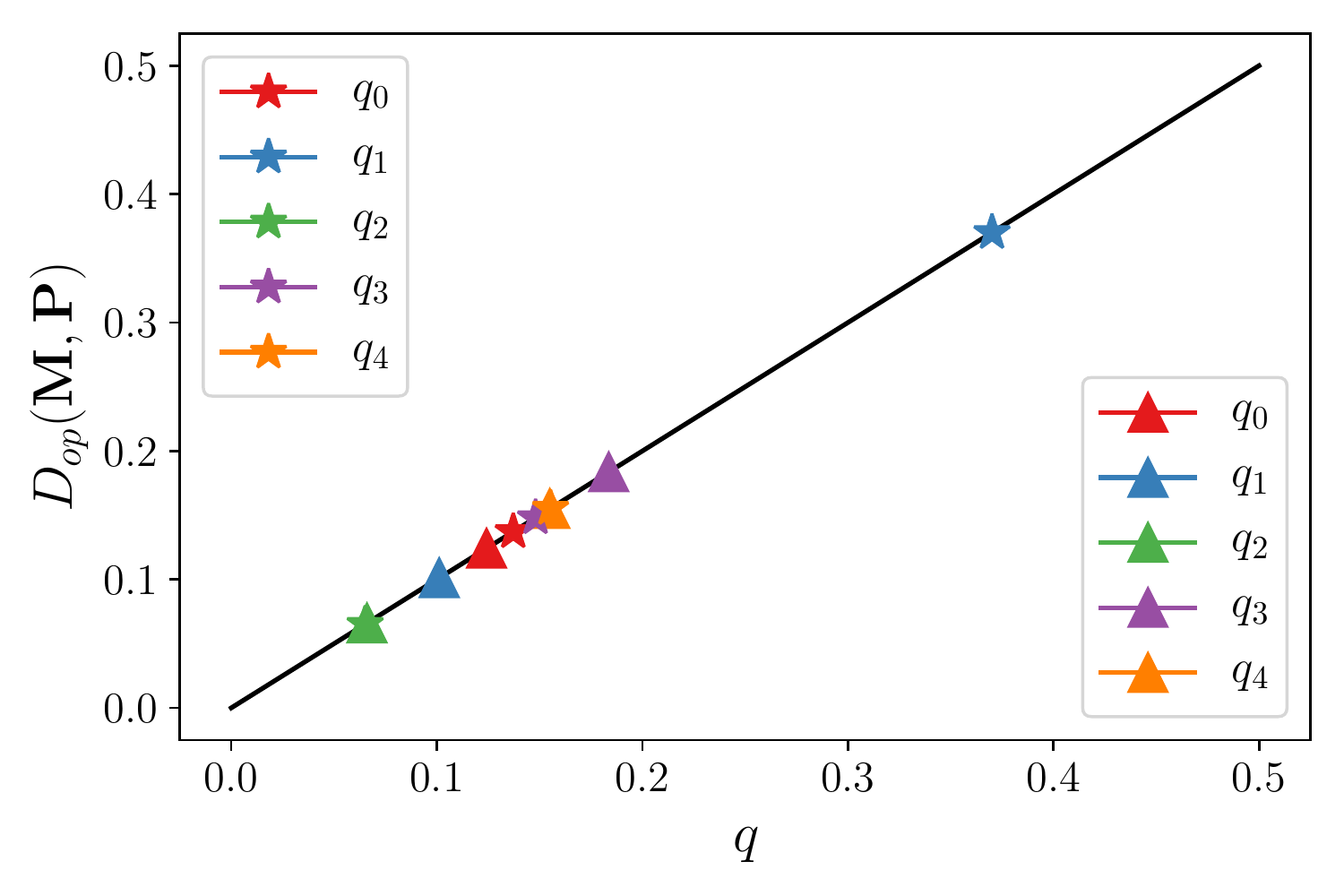}
		\caption{\label{fig:qubit_noise}
			The operational distance $D_{op}\rbracket{ \M,\P }$ between the ideal single-qubit projective measurement $\P$ and the measurement with classical noise $\M$  plotted against the probability $q$ of erroneous detection. The figure presents the regime in which $q>p$ (see Eq.~\eqref{eq:pq}), which is the typical situation in experiments. 
			The stars (triangles) correspond to the experimental values of $q$ obtained in the detector tomography experiments on IBM's (Rigetti's) device.}
	\end{figure}

	For the case of single-qubit projective measurement, without looss of generality we may write effects of a perfect detector POVM $\M^{\text{\text{ideal}}}\equiv \P$ in the computational basis as
	\begin{equation}\label{projective_qubit}
	P_1= %
	\begin{bmatrix}%
	1&0\\%
	0&0
	\end{bmatrix}\ , \
	P_2= %
	\begin{bmatrix}%
	0&0\\%
	0&1%
	\end{bmatrix}\ .
	\end{equation}
	Let us assume that the measurement that is actually implemented is of the form $\M^\mathrm{exp}\equiv\M = \Lambda \P$, for some invertible stochastic transformation $\Lambda$. This means that effects of $\M$ can be written as
	\begin{equation} \label{eq:pq}
	M_1= %
	\begin{bmatrix}%
	1-p&\quad 0\\%
	0&\quad q%
	\end{bmatrix}\ ,\ 
	M_2= %
	\begin{bmatrix}%
	p&\quad 0\\%
	0&\quad 1-q%
	\end{bmatrix}\ ,
	\end{equation}
	where $p,q \in \sbracket{0,1}$ are probabilities of erroneous detection for outcomes $1$ and $2$, respectively. 	It is now straightforward to verify that
	\begin{equation}\label{eq:example_stochastic}
	\Lambda= %
	\begin{bmatrix}%
	1-p&\quad q\\%
	p&1-q%
	\end{bmatrix}\ .
	\end{equation} 
	From $\Lambda$ we can construct the \textbf{correction matrix}
	\begin{equation}\label{example_correction}
	\Lambda^{-1}=
	\frac{1}{1-p-q} %
	\begin{bmatrix}%
	1-q&\quad -q\\%
	-p&1-p%
	\end{bmatrix}\ ,
	\end{equation}	
	which can be used in the future to correct any statistics obtained in one-qubit experiments performed on this device. 
	The adjustment simply amounts to multiplying column vector of statistics from the left by $\Lambda^{-1}$. Finally, to illustrate how error probabilities $p$ and $q$ affect the distance of $\M$ from the ideal detector $\P$, we compute the operational distance 
	\begin{equation}\label{eq:distance single}
	D_{op}\rbracket{\M,\P}=\max\cbracket{p,q}.
	\end{equation}
	The plot of this dependence is presented in Fig.~\ref{fig:qubit_noise}. 
	
	\subsubsection{Multiple qubits}
	Considerations from the previous subsection may be easily generalized for multiple-qubit systems. 
	In general, detector errors may exhibit correlations between qubits, therefore QDT should be performed via simultaneous measurements on multiple qubits. The problem of tomographic reconstruction is then exponential in the number of qubits $K$, since it requires preparation of exponentially many quantum states. However, QDT can be performed efficiently if the readout erros affecting the qubits are uncorrelated, i.e.,
	\begin{equation}
	\Lambda^{(K)}= \bigotimes
	_{i=1}^K \Lambda_i\ ,\ \Lambda_i= %
	\begin{bmatrix}%
	1-p_i&\quad q_i\\%
	p_i&1-q_i%
	\end{bmatrix}\ ,
	\end{equation}
	where $\Lambda_i$ is the stochastic matrix describing the readout noise on the $i$th qubit. If that is the case, it suffices to perform multiple single-qubit QDTs, which makes the problem linear in the number of qubits. Interestingly, the operational distance between the noisy detector  $\M^{(K)}=\Lambda^{(K)} \P^{(K)}$ and a multi-qubit projective measurement $\P^{(K)}=\otimes_{i=1}^K \P_i $ can be computed analytically (see Appendix~\ref{app::proofs} for a detailed derivation)
	\begin{equation}\label{eq:uncorrERR}
	D_{op}\rbracket{\M^{(K)},\P^{(K)}}=1-\prod_{i=1}^K\left(1-D_{op}\rbracket{\Lambda_i \P_i,\P_i}\right)\ ,
	\end{equation} 
	where $D_{op}\rbracket{\Lambda_i \P_i,\P_i}=\max\cbracket{p_i,q_i}$ is the operational distance between ideal and noisy detector on the $i$th qubit. Importantly for small individual errors (i.e. for $D_{op}\rbracket{\Lambda_i \P_i,\P_i}\ll 1 $)  implies that the total error $D_{op}\rbracket{\M^{(K)},\P^{(K)}}$ is approximately additive
	\begin{equation}
	D_{op}\rbracket{\M^{(K)},\P^{(K)}}\approx \sum_{i=1}^K D_{op}\rbracket{\Lambda_i \P_i,\P_i} , 
	\end{equation}
	This motivates the usage of our error-mitigation procedure in this regime.

	\section{Error analysis}\label{sec: Errors}

	So far we have considered idealized scenarios in which noise affecting the measurements was perfectly classical and we could repeat experiments infinitely many times. 
	Now we will provide an analysis of the possible deviations from the ideal model for the problem of reconstruction of measurement statistics. Specifically, we will describe three sources of errors:
	\begin{enumerate}
		\item Not entirely classical noise -- it might happen that the noise is not of purely classical nature (as described in Sec. \ref{sec:theory}), but also have some non-classical unitary rotation part.
		\item We have only access to a finite number of experiments, which introduces the statistical noise when we reconstruct probability distributions.
		\item Finding the closest probability vector in the case when the correction yields non-physical probability distribution.
	\end{enumerate}
	
	We note that we omit here errors resulting from the imperfect tomography of measurements, which we shall address in the future work\footnote{To account for imperfect QDT, we would need a method for estimation of confidence intervals for the operational distance between POVMs, which we have not developed yet (some progress in this regards have been recently developed for quantum states \cite{Guta2018})}.

	\subsection{Non-classical noise}
	
	Typically noise affecting quantum measurements cannot be described by  \eq{eq::classical_noise}.
	Let us consider the situation, in which the measurement that is actually being implemented is of the form 
	\begin{align}\label{eq::general_noise}
	\M^{\text{exp}}=\Lambda\M^{\text{\text{ideal}}}+\mathbf{\Delta}\ ,
	\end{align}
	where we have decomposed a POVM into a part $\Lambda\M^{\text{\text{ideal}}}$ that represents ideal POVM affected by the classical noise (as in Eq.~\eqref{eq::general_noise}) and $\mathbf{\Delta}$ which represents every other (non-classical) errors in reconstructed POVM. 
	Of course, having access to $\M^{\text{exp}}$ (e.g., from QDT) and $\M^{\text{\text{ideal}}}$ (from the theoretical model), decomposition presented in \eq{eq::general_noise} may be done in arbitrary way.
	However, in the case of projective $d$-outcome measurement in the computational basis, the following ansatz seems natural.
	Namely, we propose to consider a diagonal part of the POVM as containing information about $\Lambda$, while all off-diagonal terms should be regarded as non-classical part of the noise $\mathbf{\Delta}$ (see also Remark~\ref{rem::generic_lambda_reconstruction}).
	
	Clearly, non-zero $\mathbf{\Delta}$ affects our error-mitigation procedure.
	In the case of a noisy POVM of the form given in \eq{eq::general_noise}, it is impossible to reverse the effects of such noise solely by classical post-processing. Computing the vector of statistics for arbitrary quantum state $\rho$ with the help of Eq.~\eqref{eq::general_noise} yields (in analogy to Eq.\eqref{eq::statistics_noise})
	\begin{equation}\label{eq::statistics_noise_general}
	\vec{p}_{\text{exp}}=\Lambda \vec{p}_{\text{\text{ideal}}}+\tilde{\mathbf{\Delta}},
	\end{equation}
	where  $\tilde{\mathbf{\Delta}}$ denotes a generic disturbance of experimental statistics which arises due to the presence of non-classical part of the noise $\mathbf{\Delta}$. Its elements may be defined as $\tilde{\Delta}_i\coloneqq \Tr{\rho \Delta_i}$, where $\rho$ is an \emph{unknown} quantum state and $\Delta_i$ corresponds to the non-classical part of $i$th effect. 
	Applying $\Lambda^{-1}$ to Eq.~\eqref{eq::statistics_noise_general}
	gives
	\begin{equation}\label{general_noise_reverse}
	\Lambda^{-1} \vec{p}_{\text{exp}}= \vec{p}_{\text{\text{ideal}}}+\Lambda^{-1}\tilde{\mathbf{\Delta}}\ ,
	\end{equation}
	which clearly does not leave us with the ideal statistics, but consists also some additional term $\Lambda^{-1}\tilde{\mathbf{\Delta}}$.

	However, if non-classical part $\mathbf{\Delta}$ is small compared to the term $\Lambda \M^{\rbracket{ideal}}$, we propose to neglect the non-classical part $\mathbf{\Delta}$ and perform our error-mitigation procedure \textit{as if there were only classical noise}.
	Such action will clearly introduce some error into resulting estimated probability vectors. 
	We quantitatively characterize the error introduced by neglecting non-classical part of the noise by finding an upper bound on $D_{TV}\left(\Lambda^{-1}\vec{p}_{\text{exp}},\vec{p}_{\text{ideal}}\right)$ in terms of the operational distance between the ideal measurement $\M^{\text{exp}}$ and  $\Lambda\M^{\text{ideal}}$. Using Eq.\eqref{eq::statistics_noise_general} we obtain
	\begin{align*}
	D_{TV}\left(\Lambda^{-1}\vec{p}_{\text{exp}},\vec{p}_{\text{ideal}}\right)&=\frac{1}{2}||\Lambda^{-1} \vec{p}_{\text{exp}} - \vec{p}_{\text{ideal}}||_{1} = \\
	\frac{1}{2}||\Lambda^{-1}\tilde{\mathbf{\Delta}}||_{1} &  \leq \frac{1}{2}||\Lambda^{-1}||_{1\rightarrow 1}|| \tilde{\mathbf{\Delta}}||_{1}\ , \numberthis
	\end{align*}
	where in the inequality in the second line we have used the standard inequality $||A x||_1\leq ||A||_{1\rightarrow 1} ||x||_1$ valid for all vectors in $\mathbb{R}^n$ and linear transformations\footnote{The norm $||A||_{1\rightarrow 1}$ denotes the operator norm of $A$ understood as an operator between in $\mathbb{R}^n$ equipped with $l_1$ norm. It can be easily compute as it equals to the maximum of $l_1$ norms of columns of $A$. } $A:\mathbb{R}^n \rightarrow \mathbb{R}^n$.  
	Performing the optimization over all quantum states (note that $\tilde{\mathbf{\Delta}}$ implicitly depends on a quantum state $\rho$), and using Eq.\eqref{eq::DopDtv} finally yields the bound
	\begin{equation}\label{eq::error_diag}
	D_{TV}\left(\Lambda^{-1}\vec{p}_{\text{exp}},\vec{p}_{\text{ideal}}\right) \leq\\ ||\Lambda^{-1}||_{1\rightarrow 1} D_{op}\left(\M^{\text{exp}},\Lambda\M^{\text{ideal}}\right)
	\end{equation}
	We see that upper bound for the error that is introduced by our error-mitigation procedure, due to neglecting the non-classical part of the noise, can be expressed by $1 \rightarrow 1$ norm of $\Lambda^{-1}$ (i.e. maximal $l_1$ norm of its columns) and by the operational distance between the actually implemented measurement  $\M^{\text{exp}}$ (given by \eq{eq::general_noise}), and the part of POVM which is affected only by classical noise , i.e., $\Lambda\M^{\text{ideal}}$.

	\begin{rem}
		\label{rem::generic_lambda_reconstruction}
		The above considerations were based on the decomposition of the imperfect measurements on the 'classical' ($\Lambda\M$) and the 'non-classical' ($\vec{\Delta}$) part. 
		In general, such decomposition may be done in a somewhat arbitrary way. In principle, one could define an optimization problem, which should minimize, e.g., some suitably-chosen cost function depending on  $||\Lambda^{-1}||_{1\rightarrow 1}$ and $||\mathbf{\Delta }||_1$ so aiming to minimize the upper bound on the error of the error-mitigation procedure.
	\end{rem}

	\subsection{Statistical errors}\label{subsec:STAT}
	We will now study how finite-statistics errors affect the performance of our method.
	First of all, let us note that in experiments we do not have access to the probability vector $\vec{p}_{\text{exp}}$ which we used previously, but only to its \emph{estimator}. 
	In what follows we will focus on the natural  Maximum-Likelihood (ML) estimator for which estimates of individual probabilities $p_i$ are given by relative frequencies $n_i / N$  of events observed in the experiment repeated multiple times.
	Therefore, we shall now call the experimental statistics vector $\vec{p}_{\text{exp}}^{\text{est}}$, putting emphasis on the fact that it is an estimator.
	
	The quality of the estimation procedure can be quantified by $\mathrm{Pr}_{err}(\epsilon) \coloneqq \mathrm{Pr}\rbracket{D_{TV}\rbracket{\vec{p}_{\text{exp}}^{\text{est}},\vec{p}_{\text{exp}}}\geq \epsilon}$, i.e., the probability that the TV distance between the estimated statistics vector $\vec{p}_{\text{exp}}^{\text{est}}$ and the true vector $\vec{p}_{\text{exp}}$ is greater than some threshold  $\epsilon$.
	Hence, the value of $1-\mathrm{Pr}_{err}(\epsilon)$ may be interpreted as the acceptable confidence level for our estimation.
	Let us denote by $N$ the number of samples (experimental runs) and by $n$ the number of outcomes of the considered measurement (since we mainly focus on the standard projective measurements, $n$ coincides with the dimension of the considered Hilbert space). 
	In work \cite{Weissman2003} authors proved that in this scenario with probability at least $1- \mathrm{Pr}_{err}$
	\begin{equation}\label{eq::statistical_error}
	D_{TV}\left(\vec{p}_{\text{exp}}^{\text{est}},\vec{p}_{\text{exp}}\right) \leq \sqrt{\frac{\log\rbracket{2^n-2}-\log\rbracket{\mathrm{Pr}_{err}}}{2N}}\eqqcolon \epsilon\ .
	\end{equation}
	The above inequality gives us the upper bound for the TV-distance between estimated statistics (frequencies) vector $\vec{p}^{\text{est}}_{\text{exp}}$ and actual probability vector $\vec{p}_{\text{exp}}$ as a function of the number of experimental runs and the accepted error probability.
	
	Let us finally stress that we are not interested in the statistical error itself as it is inherently present in any estimation scheme, whenever we have on our disposal only a finite number of samples. 
	Instead, we want to understand how finite-sample statistics affect our procedure. Specifically, we want to bound the TV distance of our corrected estimated statistics $\Lambda^{-1}\vec{p}_{\text{exp}}^{\text{est}}$ to the ideal probability vector $\vec{p}_{\text{ideal}}$. This can be easily done in a manner analogous to the derivation leading to  \eq{eq::error_diag}. 
	The final bound is the following
	\begin{align*}\label{eq::error_diag_stat}
	&D_{TV}\rbracket{\Lambda^{-1}\vec{p}_{\text{exp}}^{\text{est}},\vec{p}_{\text{ideal}}}\leq\\& 
	D_{TV}\rbracket{\Lambda^{-1}\vec{p}_{\text{exp}}^{\text{est}},\Lambda^{-1}\vec{p}_{\text{exp}}} + D_{TV}\rbracket{\Lambda^{-1}\vec{p}_{\text{exp}},\vec{p}_{\text{ideal}}}\leq\\&
	||\Lambda^{-1}||_{1\rightarrow 1} \epsilon+||\Lambda^{-1}||_{1\rightarrow 1} D_{op}\rbracket{\M^{\text{exp}},\Lambda\M^{\text{ideal}}} \eqqcolon \delta\ ,
	\numberthis 
	\end{align*}
	where the first inequality is a consequence of the triangle inequality, and the derivation of the second inequality is fully analogous to that of Eq.~\eqref{eq::error_diag}.
	In the upper bound above, the first term in the sum accounts for the statistical errors while the second term accounts for the non-classical part of the noise.

	\subsection{Non-physical probability vectors}\label{sec: Errors_c}
	Due to the coherent and statistical errors described above, it may happen that the corrected vector of the estimated statistics $\Lambda^{-1}\vec{p}^{\text{est}}_{\text{exp}}$ will not be a proper probability vector. Generally, the vector obtained in our procedure may contain negative elements that will, however, sum up to 1  (this is because the inverse of left-stochastic matrix still has columns of which elements sum up to 1). When this is the case, we propose to find the \textit{corrected} vector $\vec{p}_{\text{exp}}^{_\ast}$ which is the closest probability vector to $\Lambda^{-1}\vec{p}^{\text{est}}_{\text{exp}}$ in Euclidean norm\footnote{We note that the choice of this norm in the error-mitigation procedure is somewhat arbitrary. We have chosen the Euclidean for sake of the mathematical convenience (the relevant optimization problem can be easily solved).}. 
	In other words, if obtained vector $\Lambda^{-1}\vec{p}^{\text{est}}_{\text{exp}}$ is not a proper probability vector, we use the \emph{corrected} vector $\vec{p}_{\text{exp}}^{_\ast}$ th5at is the solution to the following problem 
	\begin{equation}\label{eq::est_problem}
	\vec{p}^{_\ast}_{\text{exp}}=\underset{\forall_i p_{i}\geq 0,\ \sum_{i=1}^{n} p_{i}=1}{\mathrm{argmin}} \left(||\Lambda^{-1}\vec{p}^{\text{est}}_{\text{exp}}-\vec{p}||_2 \right)\ ,
	\end{equation} 
	where $||\cdot ||_2$ denotes the Eucledian norm. 
	The above problem can be easily solved by convex optimization solvers like \textit{cvxopt} \cite{cvxopt}.
	Furthermore, the error $\alpha$ introduced by this method may be quantified in terms of TV-distance between new corrected vector $\vec{p}^{_\ast}_{\text{exp}}$ and nonphysical $\Lambda^{-1}\vec{p}^{\text{est}}_{\text{exp}}$
	\begin{align}\label{eq::error_strange}
	\alpha\coloneqq \frac{1}{2} ||\vec{p}^{_\ast}_{\text{exp}}-\Lambda^{-1}\vec{p}_{\text{exp}}^{\text{est}}||_1.
	\end{align}
	
	In order to account for the overall error of our procedure, we modify the \eq{eq::error_diag_stat} by putting $\vec{p}^{_\ast}_{\text{exp}}$ in place of $\Lambda^{-1}\vec{p}_{\text{exp}}^{\text{est}}$. The usage of the triangle inequality gives the following upper bound for the overall error in our procedure 
	\begin{equation}\label{eq:error_overall}
	D_{TV}\rbracket{\vec{p}^{_\ast}_{\text{exp}},\vec{p}_{\text{ideal}}} \leq \delta+\alpha \ ,
	\end{equation}
	where $\delta$ and $\alpha$ are defined in Eq.~\eqref{eq::error_diag_stat} and Eq.~\eqref{eq::error_strange}, respectively. 
	The quantity $\delta+\alpha$ can be considered as the overall error that our procedure yields for the problem of estimation of the probability vector $\vec{p}_{\text{ideal}}$. This upper bound on $D_{TV}\rbracket{\vec{p}^{_\ast}_{\text{exp}},\vec{p}_{\text{ideal}}}$ is a function of four variables: the number of experimental runs $N$ used in the procedure of probability estimation, accepted error probability $\mathrm{Pr}_{err}$, the reconstructed measurement $\M^{\text{exp}}$, and the vector of statistics obtained in a given experiment $\vec{p}_{\text{exp}}^{\text{est}}$.

	\subsection{When is the mitigation successful?}
	\begin{figure}[t]
		%		\begin{subfigure}
		%			\centering
		%			\includegraphics[scale=0.5]{fig_1q_delta_real.pdf}
		%			\label{fig::vis_1q_delta_a}
		%		\end{subfigure}
		%		\begin{subfigure}
		%			\centering
		%			\includegraphics[scale=0.5]{fig_1q_delta_real_rigetti.pdf}
		%			\label{fig::vis_1q_delta_b}
		%		\end{subfigure}
		\centering
		\includegraphics[scale=0.5]{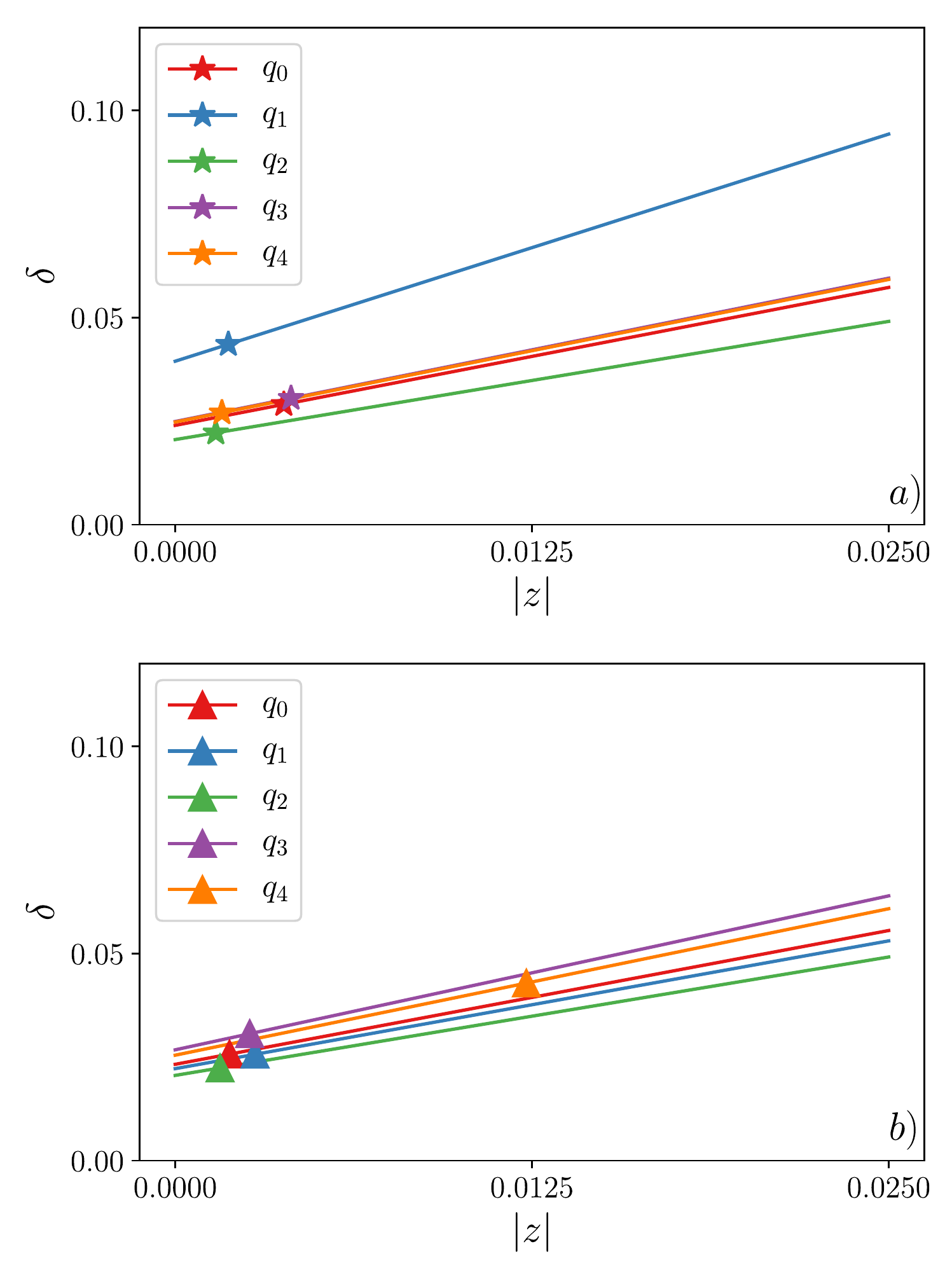}
		
		\caption{ \label{fig::vis_1q_delta}
			Dependence of the error upper bound $\delta$  on the magnitude of the coherent error $|z|$ for noisy single-qubit measurements. We depict the interval of relevant $|z|$ values obtained from the detector tomography performed on a) IBM's, b) Rigetti's devices.  Stars and triangles correspond to the actual experimental values of $|z|$. Statistical error is fixed at value $\epsilon=0.018$, which corresponds to setting  $\mathrm{Pr}_{err}=0.01$ when $N=8192$.}
	\end{figure}
	
	\begin{figure*}
		\includegraphics[scale=0.14]{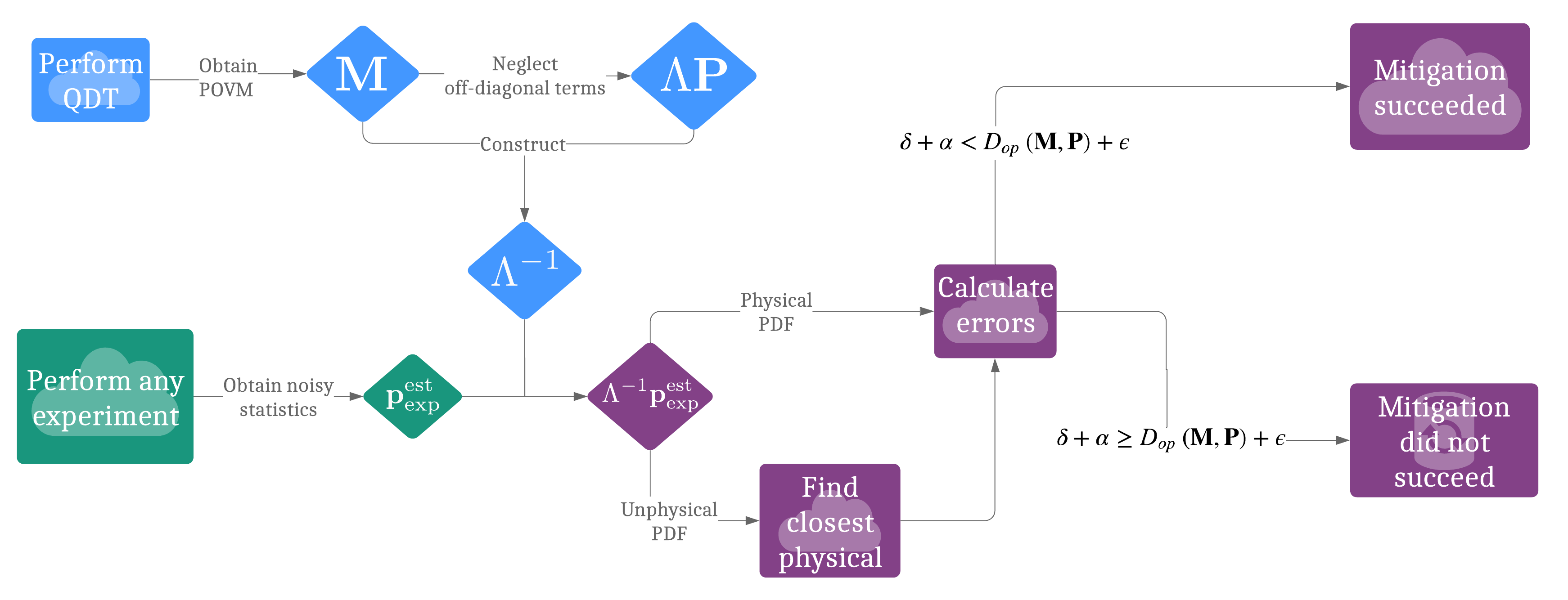}
		\caption{\label{fig::correction_flowchart}Pictorial representation of the mitigation scheme for the case of projective measurements. 
			Blue boxes correspond to the pre-processing stage.
			In this stage, it is first necessary to perform the tomography of the noisy detector. This gives access to the classical description of the noisy POVM $\M^{\text{exp}}$. Neglecting the off-diagonal terms of its effects gives the classical part of the readout noise $\Lambda$, from which one obtains the correction matrix $\Lambda^{-1}$.	
			Green boxes correspond to performing an arbitrary experiment on the noisy device. 
			The error-mitigation scheme itself is presented by the purple boxes.
			Upon correcting experimental statistics $\vec{p}^\mathrm{est}_{\text{exp}}$, one needs to determine if $\Lambda^{-1}\vec{p}^\mathrm{est}_{\text{exp}}$ is a physical probability vector. 
			If not, one performs the additional step of finding the physical probability vector closest to $\Lambda^{-1}\vec{p}^\mathrm{est}_{\text{exp}}$ . 
			The final step is to calculate the upper bound on the error $D_{TV}\left(\vec{p}^{_\ast}_\mathrm{exp},\vec{p}_\mathrm{ideal}\right)$ given by  $\delta+\alpha$ (Eq.~\eqref{eq:error_overall}).  The decision whether such a mitigation is successful or not is based on the comparison of this bound with  $D_{op}\rbracket{\P,\M} +\epsilon$. }
	\end{figure*}

	We would like to address now a crucial question. Namely, when can we consider our error-mitigation procedure to be successful?
	The answer to this can be given by analyzing the upper bound $\delta+\alpha$ on the error of estimation of probability vector that can be potentially obtained by the error-mitigation scheme (resulting from the presence of coherent errors and finite-statistics errors).
	
	Imagine that we have on our disposal only the noisy quantum measurement $\M^{\text{exp}}$ that produces $N$ samples from the distribution $\vec{p}_{\text{exp}}$.
	Let $\vec{p}^{\text{est}}_{\text{exp}}$ denote the empirical estimator of this probability vector.
	By the virtue of considerations given in earlier subsections, the error resulting from imperfect measurement and the finite number of samples (for the assumed error probability $\mathrm{Pr}_{err}$)  can be upper bounded as follows
	\begin{equation}\label{eq:statNONperf}
	D_{TV}\rbracket{\vec{p}^{\text{est}}_{\text{exp}},\vec{p}^{\text{ideal}}}\leq D_{op}\rbracket{\M^{\text{exp}},\M^{\text{ideal}}} + \epsilon 
	\end{equation}
	We then propose to consider our mitigation successful if the upper bound in Eq.\eqref{eq:error_overall}, on the error $D_{TV}\rbracket{\vec{p}^{_\ast}_{\text{exp}},\vec{p}^{\text{ideal}}}$ that can be introduced by the error-mitigation procedure, is smaller then the upper bound on $D_{TV}\rbracket{\vec{p}^{\text{est}}_{\text{exp}},\vec{p}^{\text{ideal}}}$ discussed above (for the same number of samples $N$ and error probability $\mathrm{Pr}_{err}$). 
	In other words we propose the following rule
	\begin{equation}\label{eq:RULEcorr}
	\delta+\alpha < D_{op}\rbracket{\M^{\text{exp}},\M^{\text{ideal}}} + \epsilon \Rightarrow \text{mitigation succesful}\ .
	\end{equation}
	
	\begin{rem}
		Of course, the actual distance from the perfect probability distribution after post-processing is highly dependent on the quantum state that is measured. Importantly, it might happen that though inequality in~\eqref{eq:RULEcorr} holds, the particular quantum state will give statistics that after post-processing are \textit{worse} than without it in terms of the distance from the perfect probability distribution. We analyze this problem in some detail in Appendix~\ref{app::to_correct_or_not_to_correct} where we provide more support of the heuristic procedure presented above.
	\end{rem}
	
	Let us illustrate the above ideas on the simple example of single-qubit projective measurement.
	We assume that in detector tomography, noisy POVM describing our device is given by
	\begin{equation}
	M_1= %
	\begin{bmatrix}%
	1-p&\quad z\\%
	\bar{z}&\quad q%
	\end{bmatrix}\ ,\ M_2= 
	\begin{bmatrix}%
	p&\quad -z\\%
	-\bar{z}&\quad 1-q%
	\end{bmatrix},
	\end{equation}
	Unlike the case of purely classical noise discussed in Section~\ref{sec: Main}, there are also off-diagonal terms which represent non-classical part of the noise.
	If the their magnitude is much smaller than that of the diagonal elements, we may approximate our POVM by $\Lambda \P$ (where the appropiate stochastic matrix is given in Eq.~\eqref{eq:example_stochastic}) at the cost of introducing some additional errors. According to the analysis presented earlier, in order to bound this error we need to compute  $||\Lambda^{-1}||_{1\rightarrow 1}$ and $D_{op}\rbracket{\M,\Lambda\P}$. Straightforward computation give us the following expressions  
	\begin{equation}
	||\Lambda^{-1}||_{1\rightarrow 1}=\frac{1+|p-q|}{|p+q-1|}\ , D_{op}\rbracket{\M,\Lambda\P}=|z|\ . 
	\end{equation}
	Inserting these expressions into \eq{eq::error_diag_stat} we obtain the following upper bound on the error of estimation of the probability distribution with the usage of our error mitigation scheme
	\begin{equation}\label{eq:appliedBOUND} 
	D_{TV}\rbracket{\Lambda^{-1}\vec{p}_{\text{exp}}^{\text{est}},\vec{p}_{\text{ideal}}}\leq \delta=\frac{1+|p-q|}{|p+q-1|}\rbracket{|z|+ \epsilon}\ ,
	\end{equation}
	where statistical error $\epsilon$ is determined by Eq. \eqref{eq::statistical_error} and depends on the assumed level of confidence and the number of samples $N$ used in the estimation procedure.  
	For example, if we set $N=8192$, which is maximal number of experimental runs for a single quantum circuit in IBM Q devices, and $\mathrm{Pr}_{err}=0.01$, we obtain $\epsilon\approx 0.018$.  Importantly in \eq{eq:appliedBOUND} we have assumed that the estimated probability vector $\Lambda^{-1}\vec{p}_{\text{exp}}^{\text{est}}$ is a probability distribution add hence we could set $\alpha=0$ in \eq{eq::error_strange}. The visualization of overall error quantifier $\delta$ is shown in Fig.~\ref{fig::vis_1q_delta}.
	\section{Summary of the method}\label{sec::Summary}
	
	In the following, we aim to correct the output statistics obtained from a noisy measurement. 
	The error-mitigation procedure applied to projective measurements is realized by performing the following sequence of practical steps, the graphical illustration of which is given in Fig~\ref{fig::correction_flowchart}. 
	\begin{enumerate} 	
		\item Perform Quantum Detector Tomography (QDT) experiments.
		The estimation of the detector's POVM $\M$ can be done using, e.g., an algorithm which converges to the Maximum Likelihood Estimation, like the one of \cite{Fiurasek2001}. 
		\item Neglect all the off-diagonal terms of the elements of $\M$, obtaining $\Lambda\P$, where $\P$ is the ideal projective measurement we want to implement.
		\item	Reverse the stochastic matrix $\Lambda$ defined in \eq{eq::classical_noise} to obtain the "correction matrix" $\Lambda^{-1}$.
		\item After performing arbitrary experiment on the device characterized in the above way, multiply the vector of obtained frequencies $\vec{p}_{\text{exp}}^{\mathrm{est}}$ by $\Lambda^{-1}$.
		\item Check if the $\Lambda^{-1} \vec{p}_{\text{exp}}$ is a proper probability vector, i.e., its elements are positive and sum up to 1.
		\begin{itemize}
			\item \textit{If yes}, set $p^{_\ast}_\mathrm{exp}=\Lambda^{-1} \vec{p}^{\mathrm{est}}_{\text{exp}}$ and proceed to the next step.
			\item \textit{If no}, solve the problem formulated in Eq~\eqref{eq::est_problem}, obtaining a vector of corrected statistics $\vec{p}^{_\ast}_\mathrm{exp}$.
		\end{itemize}		
		\item Calculate $\delta +\alpha$, i.e., the upper bound on the total error magnitude $D_{TV}\left(\vec{p}^{_\ast}_\mathrm{exp},\vec{p}_\mathrm{ideal}\right))$. 
		Compare this with  $D_{op}\rbracket{\P,\M}+\epsilon$, i.e., with the upper bound on the error occurring without correction (Eq.~\ref{eq:statNONperf}).
		\begin{itemize}
			\item If $\delta+\alpha\geq D_{op}\rbracket{\P,\M}+\epsilon$, then the error-mitigation is considered not successful.
			\item If $\delta+\alpha< D_{op}\rbracket{\P,\M}+\epsilon$, then the error-mitigation is considered successful.
		\end{itemize}		
	\end{enumerate}
	
	\begin{rem}
		We note that if we wanted to mitigate the readout errors in the case of arbitrary ideal measurement $\M^{\mathrm{ideal}}$, we would have to modify the second step of our method. Specifically, we would have to propose a decomposition of the noise on the classical ($\Lambda\M^{\mathrm{ideal}}$) and non-classical ($\vec{\Delta}$) part (see discussion in Remark~\ref{rem::generic_lambda_reconstruction}).
	\end{rem}

	\section{Device characterization}\label{sec: Validation}

\begin{figure}[h!]
		\begin{center}
			\includegraphics[scale=0.65]{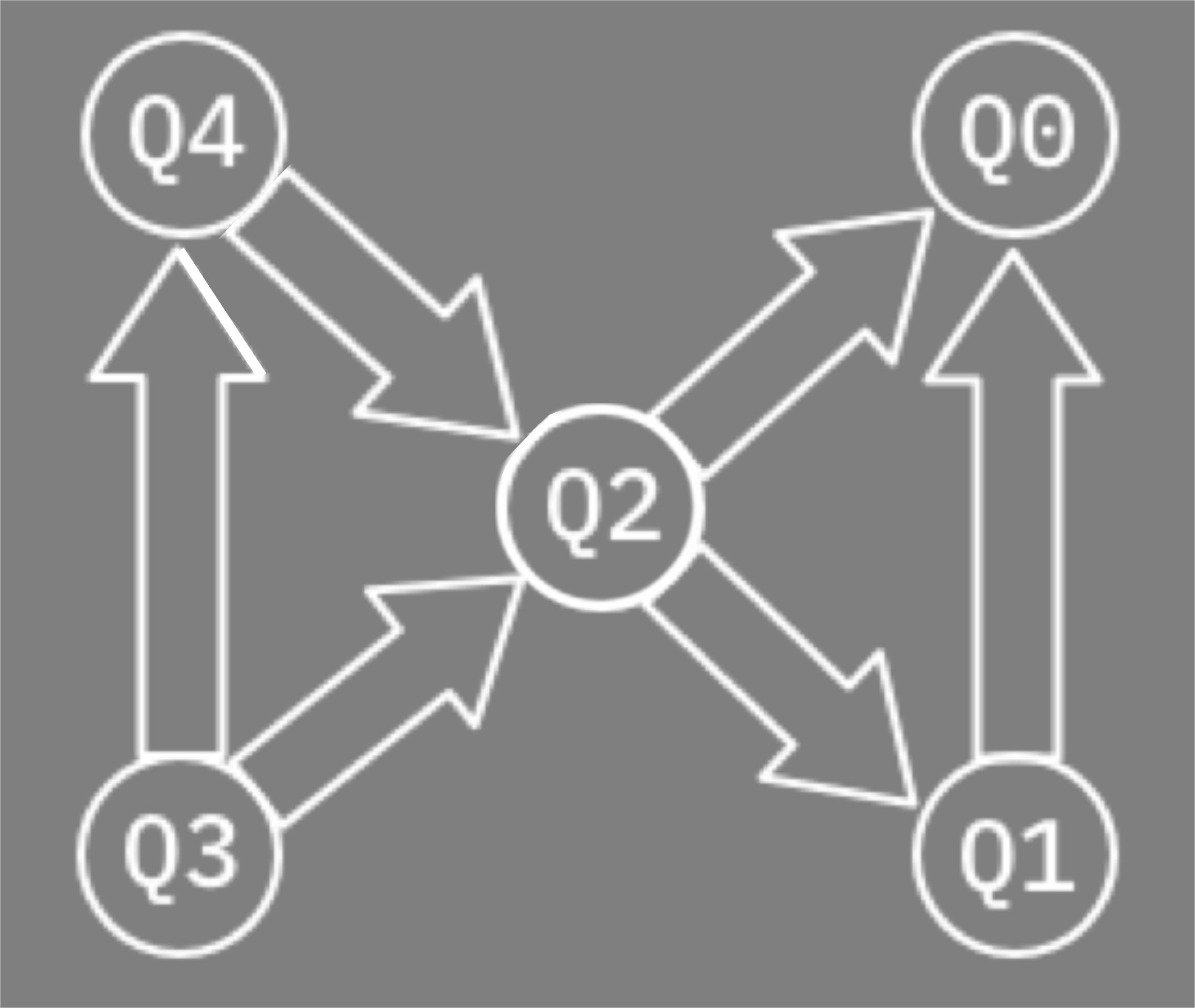}
		\end{center}
		\caption{\label{ibmqx4_scheme} The scheme of connectivity for the qubits in the \textit{ibmqx4} backend. 
			The beginning (end) of the arrows denote the control (target) qubits of the directly accessible physical CNOT gates. Source: \cite{qiskit_github}.}	
	\end{figure}

	In this section, we present experimental results\footnote{Dates of execution of all presented experiments are given in Appendix~\ref{app::exp_data}.} that confirm the (approximate) validity of the physical assumptions stated in Section \ref{sec: Main} in the IBM's device. 
	These assumptions have to be satisfied with our error-mitigation procedure to work.
	We first present the randomized benchmarking data for IBM's devices, which shows that single-qubit gates errors are small and make Quantum Detector Tomography feasible.
	Then the results of QDTs performed on IBM devices are provided.
	This shows that readout errors in these systems are indeed mostly of classical nature (see Section \ref{sec:theory}). 
	We observe that the readout errors between physically connected pairs of qubits are mostly uncorrelated  (see Fig. \ref{ibmqx4_scheme} for visual presentation of the connectivity of \textit{ibmqx4} device). 
	Furthermore, throughout the section, we present analogous data for exemplary five qubits in Rigetti's 16-qubit device \textit{Aspen-4-16Q-A}. 
	However, in the case of this device, the assumption of perfect detector reconstruction is significantly violated due to the high infidelities of the single-qubit quantum gates.

	\begin{rem}\label{rem_time_fluctuations} The experimental results presented in this work were obtained on publicly available devices. For this reason, it was not possible to perform all the experiments during a single calibration period. 
		Consequently, the data presented throughout the paper may consist of values obtained in experiments performed on different days. Importantly, to correct statistics in all experiments performed in a given calibration period, we have used the data from QDT  specific to that period. We note that  results of QDT were varying across different calibration periods. However, those fluctuations do not change qualitative, nor quantitative conclusions derived throughout our work.
	\end{rem}
	
	\begin{figure*}[]
		\centering
		\includegraphics[scale=0.585]{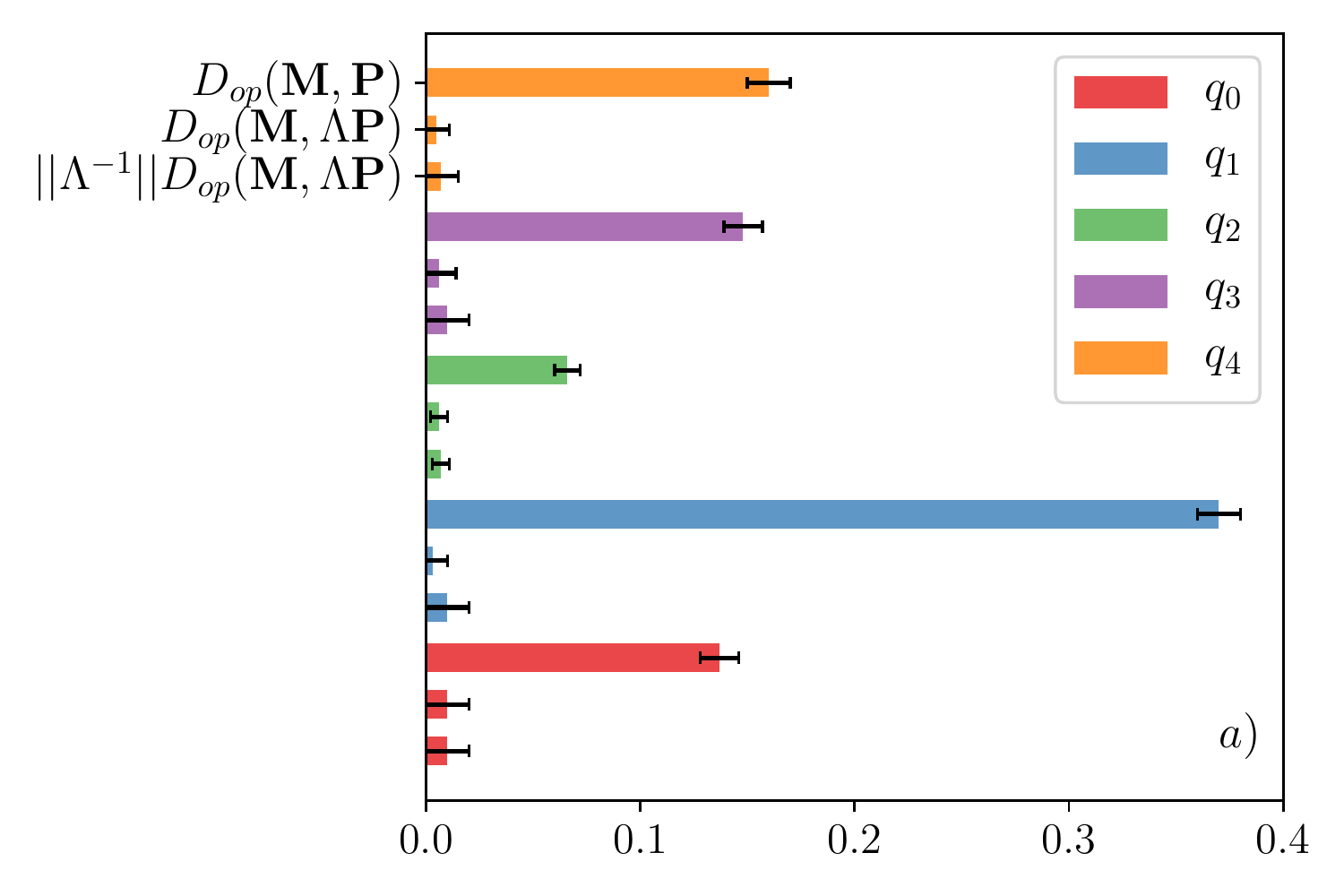}
		\label{fig::QDT_1q_a}			
		\includegraphics[scale=0.585]{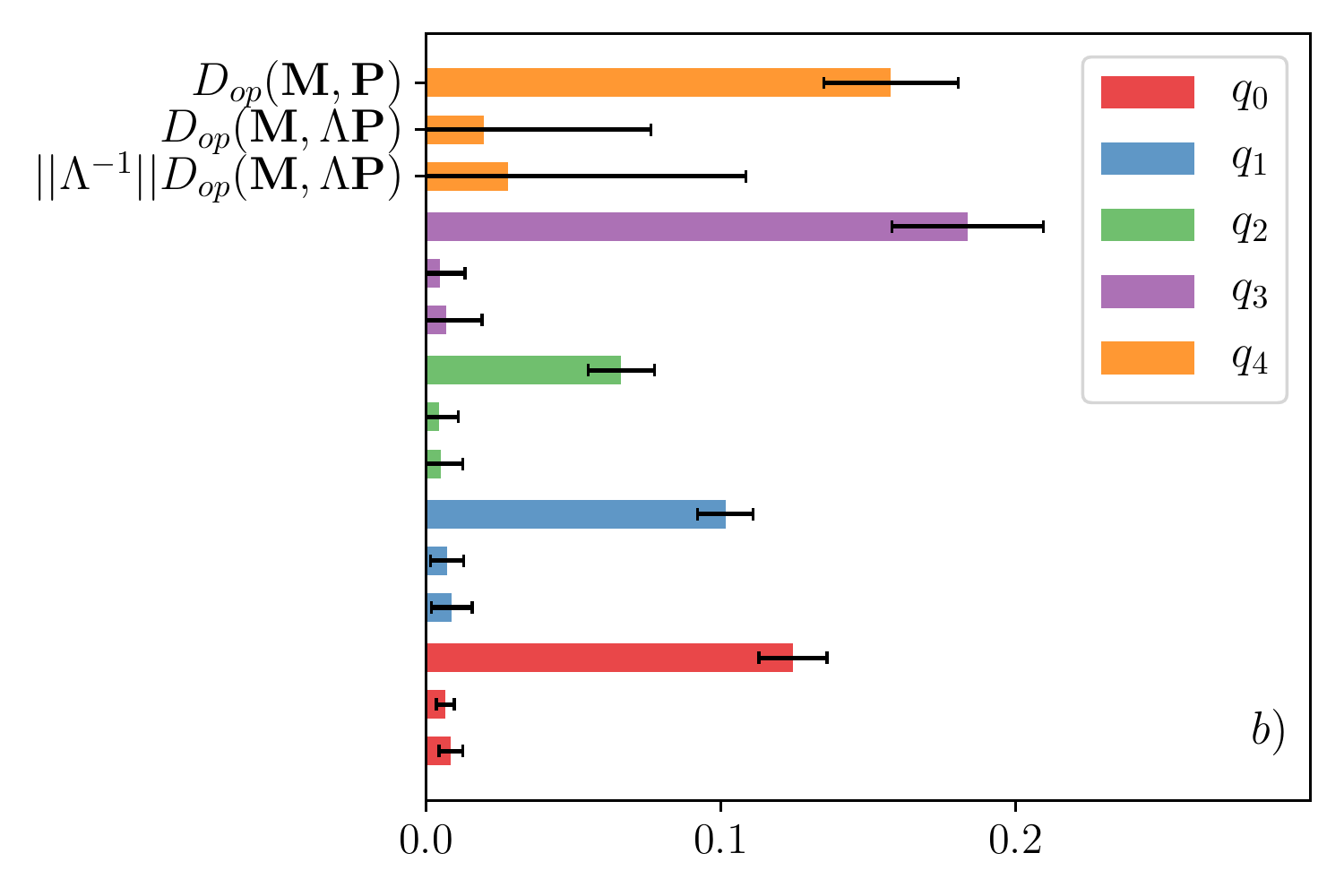}
		\label{fig::QDT_1q_b}
		\caption{ \label{fig::QDT_1q}
			Characterization of single-qubit measurements on a) IBM's \textit{ibmqx4} and b) Rigetti's \textit{Aspen-4-16Q-A}. 
			For each qubit, the first bar corresponds to the operational distance between the perfect projective measurement and the single-qubit POVM reconstructed via QDT, denoted by $D_{op}\rbracket{\M,\P}$.
			The second bar corresponds to the distance between the reconstructed POVM and the POVM obtained by neglecting its off-diagonal terms, denoted as $D_{op}\rbracket{\M,\Lambda\P}$. 
			The third bar corresponds to the upper bound for the overall error $\delta$ of our error-mitigation procedure, which does not include statistical errors.
			Each tomography experiment consisted of 6 quantum circuits: the preparation of all Pauli eigenstates.
			Each circuit was implemented $8192$ times.
			Furthermore, each QDT experiment was repeated 4 times, in order to estimate the standard deviations, the corresponding $3\sigma$ bars are shown on the plot. \hspace{20cm}	}
	\end{figure*}

	\begin{figure*}[]
		\centering
		\includegraphics[scale=0.58]{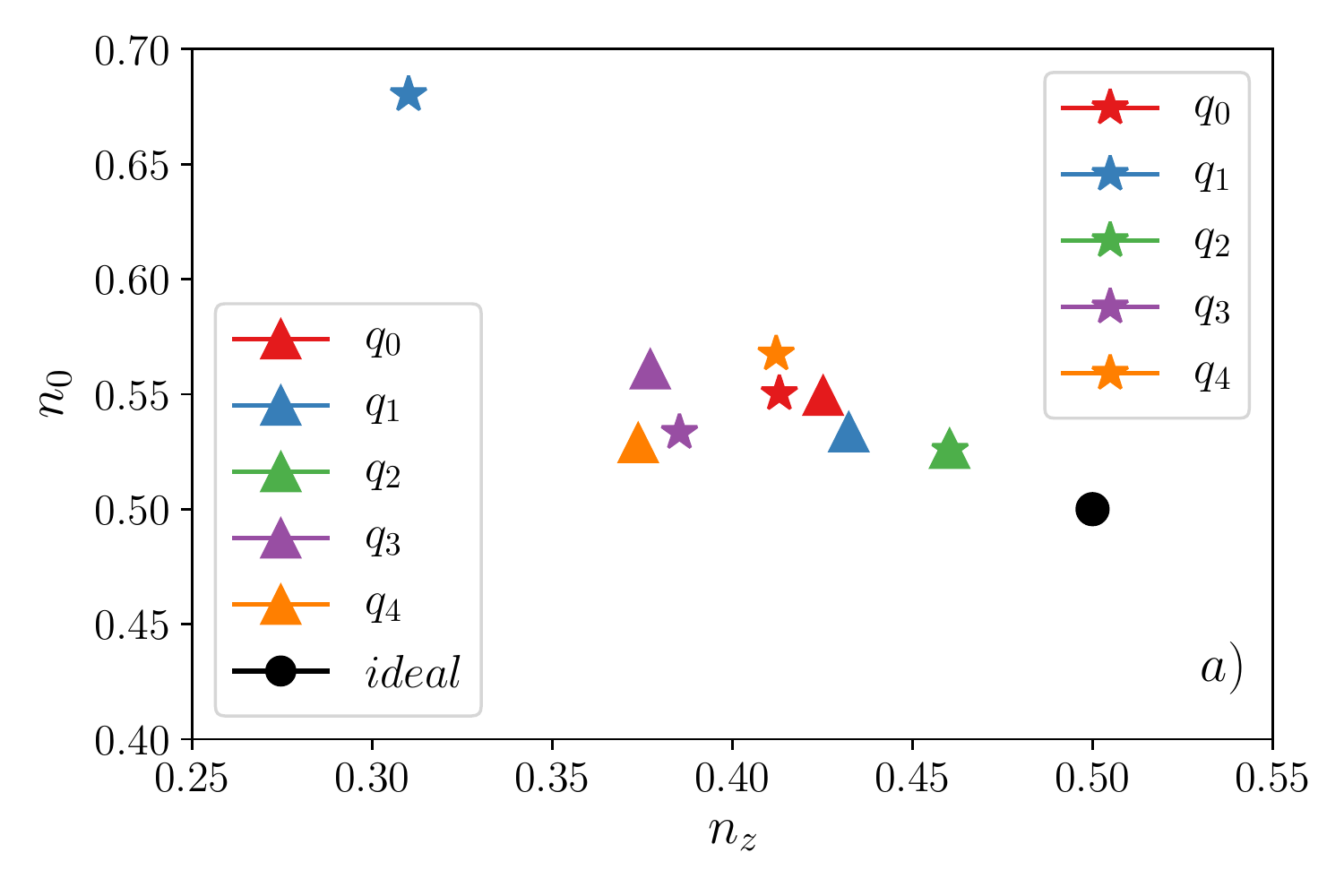}
		\label{fig::1q_arrows_a}
		\includegraphics[scale=0.58]{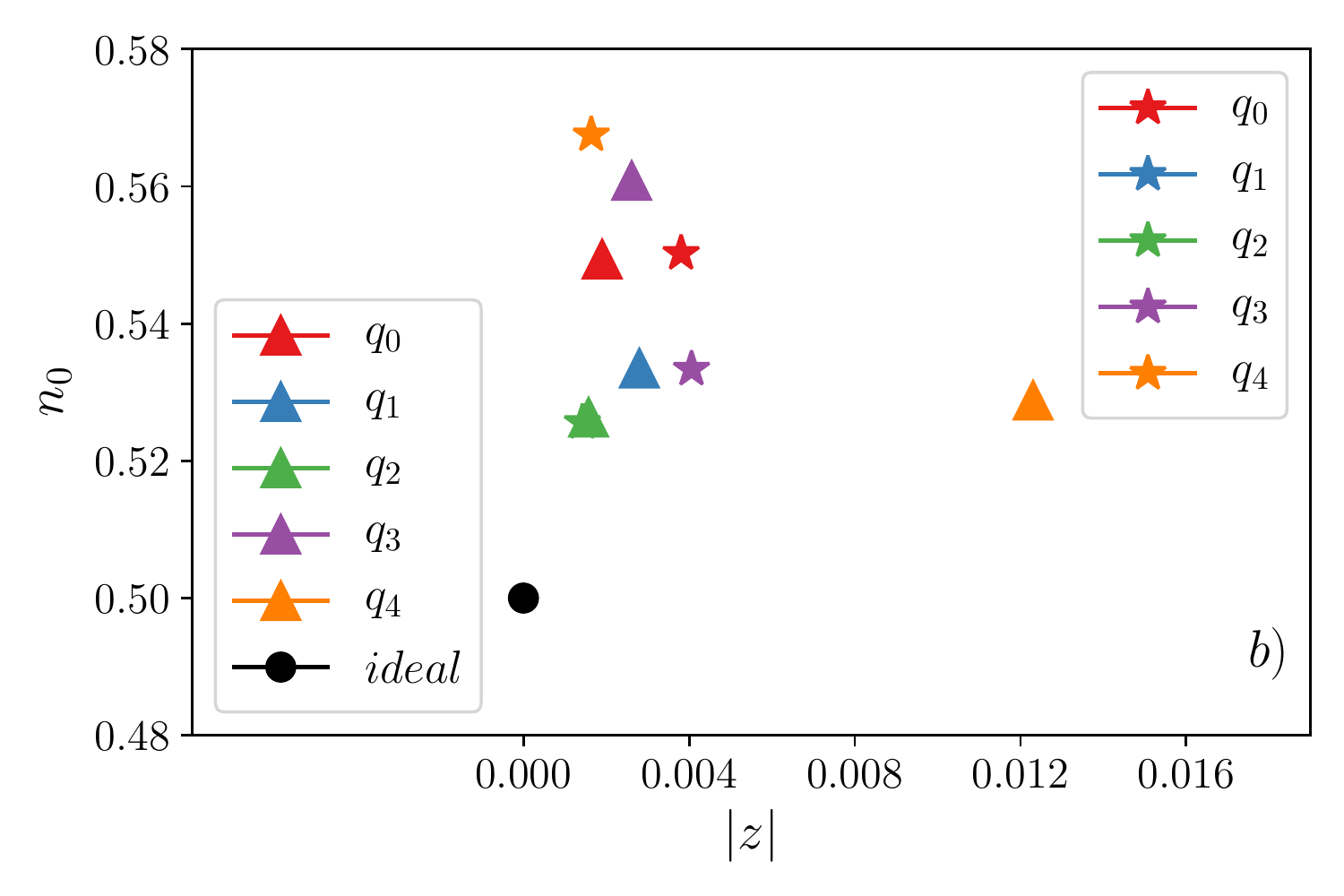}
		\label{fig::1q_arrows_b}
		%	\includegraphics[scale=0.5]{fig_arrows_nz_n0_rigetti.pdf}
		%	\label{fig::1q_arrows_c}
		%	\includegraphics[scale=0.5]{fig_arrows_n0_z_rigetti.pdf}
		%	\label{fig::1q_arrows_d}
		\caption{ \label{fig::1q_arrows}
			Visualization of single-qubit detectors for IBM's \textit{ibmqx4} (stars) and Rigetti's \textit{Aspen-4-16Q-A} (triangles). 
			We parametrize the first effect of a measurement as $M_1=\sum_{k\in\lbrace{0,x,y,z\rbrace}} n_k\sigma_k$, for $\sigma_0\coloneqq \iden$. We set $|z|\coloneqq \sqrt{n_x^2+n_y^2}$ to be the magnitude of coherent errors. In the figures, we plot the two-dimensional projections of the four-dimensional parameter space of $M_1$:  part a)  depicts the identity ($n_0$) and the $\sigma_z$ ($n_z$) components, whereas part b) shows 
			the identity ($n_0$) and the coherent ($|z|$) components.}
	\end{figure*}

	\subsection{Perfect state preparation}
	In order to be able to perform reliable Quantum Detector Tomography, one needs to assure accurate state preparation. 
	Since the tensor product of single-qubit quantum gates suffices to prepare quantum states spanning the whole operator space, it is enough to analyze only single-qubit gate errors. 
	Randomized benchmarking experiments \cite{Knill2008,Gambetta2012}  allow for inferring such errors in a manner approximately independent on state-preparation and measurement errors. 
	\begin{table}[h]	
		\begin{center}
			\includegraphics[scale=0.11]{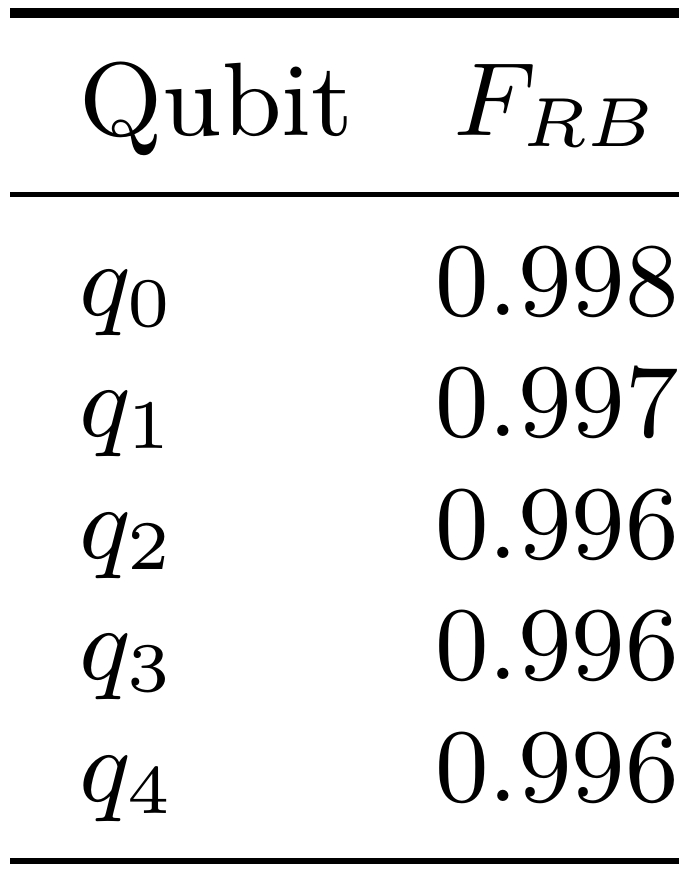} 
			\hspace{2cm}
			\label{fig::RB_a}
			\includegraphics[scale=0.16]{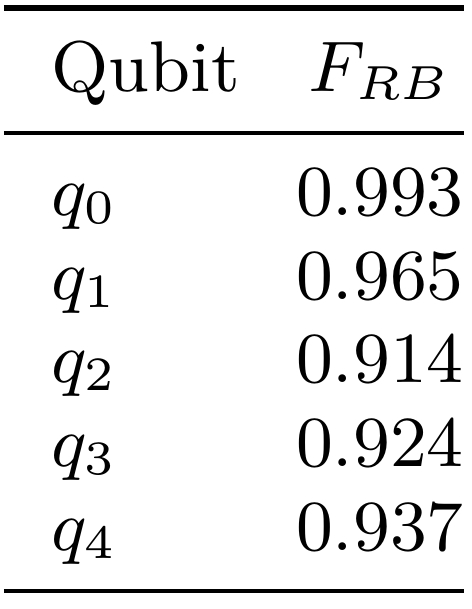}
			\label{fig::RB_b}
			\\	a) \hspace{3.6cm} b)	
		\end{center}
		\caption{\label{RB_Fig.}
			Average single-qubit gate fidelities for a) \textit{ibmqx4}, and b) \textit{Aspen-4-16Q-A}. 
			Data obtained via a) qiskit \cite{qiskit_ref} and b) Forest \cite{ref_rigetti}.
		}
	\end{table}
	In Table \ref{RB_Fig.} we list average gate fidelities $F_{RB}$ for \textit{ibmqx4} device. 
	The errors are relatively small (i.e., $1-F_{RB}$ is of order $0.1\%$) which indicates that the assumption of perfect state preparation is reasonable. 
	The average gate fidelities in Rigetti's device are, however, worse, which makes it difficult to precisely gauge the errors introduced by our error-mitigation procedure.
	
	We would like to remark that in general the problem of rigorously separating measurement, gate, and state-preparation errors is a difficult task (for recent progress in this direction, see \cite{Kohout2013,SelfCons2013,Pusey2017}). Without performing this division, we cannot precisely estimate the accuracy of our error-mitigation procedure. Let us emphasize, however, that for many practical applications (like variational quantum algorithms for example \cite{QAOA2014}) one can simply implement our scheme and test if it works in practice. 
	
	\begin{figure*}
		
		\includegraphics[scale=0.6]{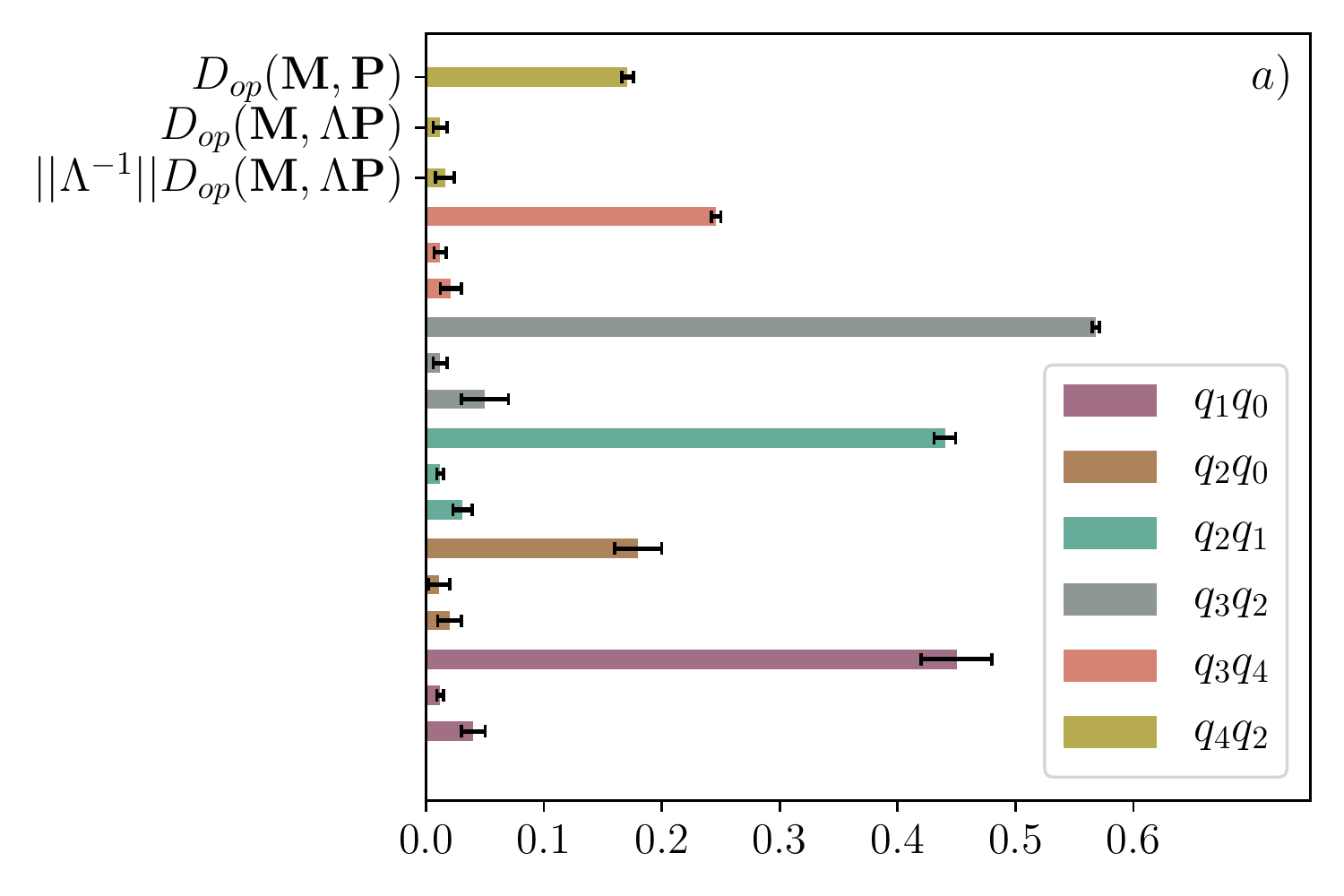}
		\label{fig::QDT_2q_a}		
		\includegraphics[scale=0.6]{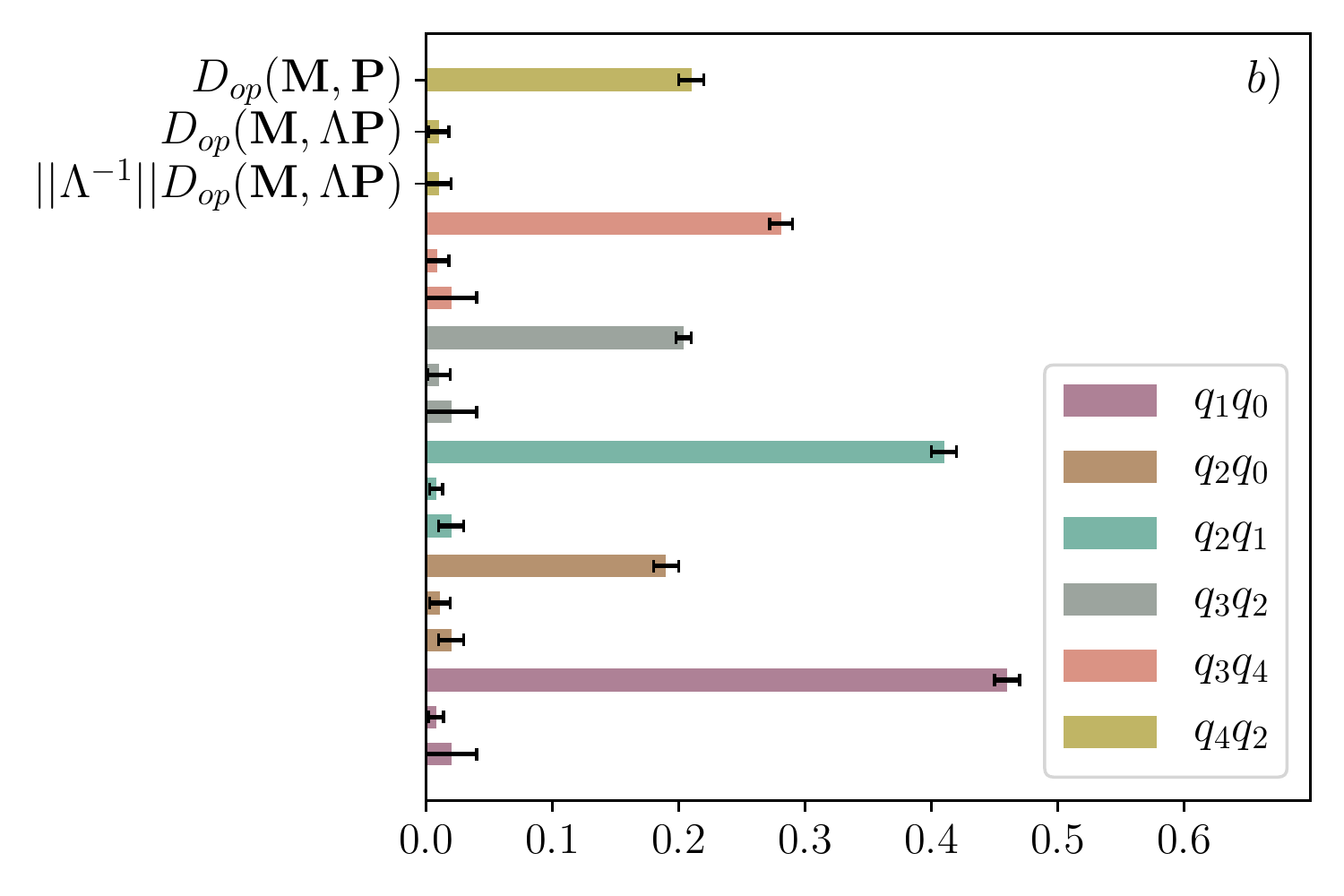}
		\label{fig::QDT_2q_b}
		\includegraphics[scale=0.6]{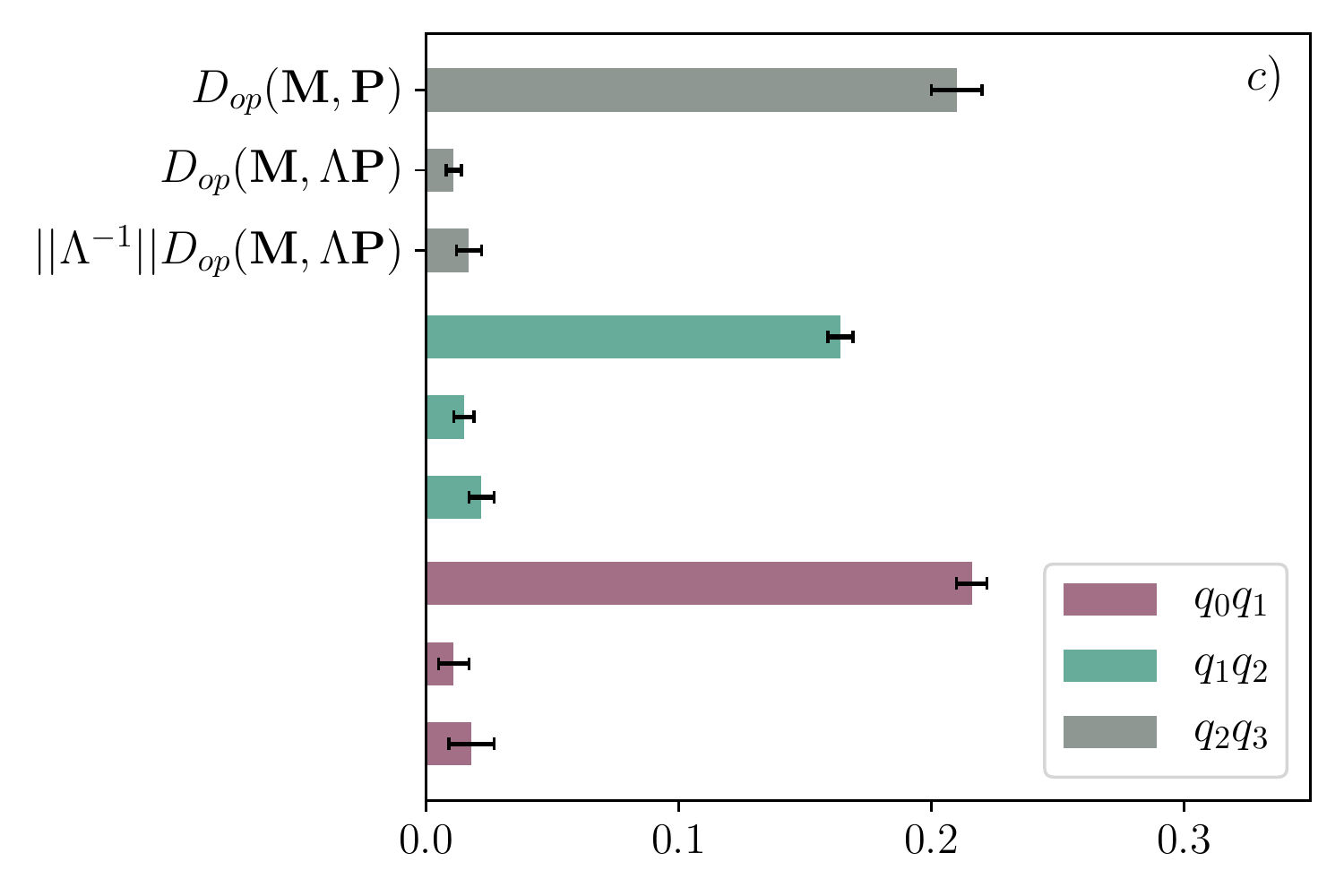}
		\label{fig::QDT_2q_a_rigetti}	
		\includegraphics[scale=0.6]{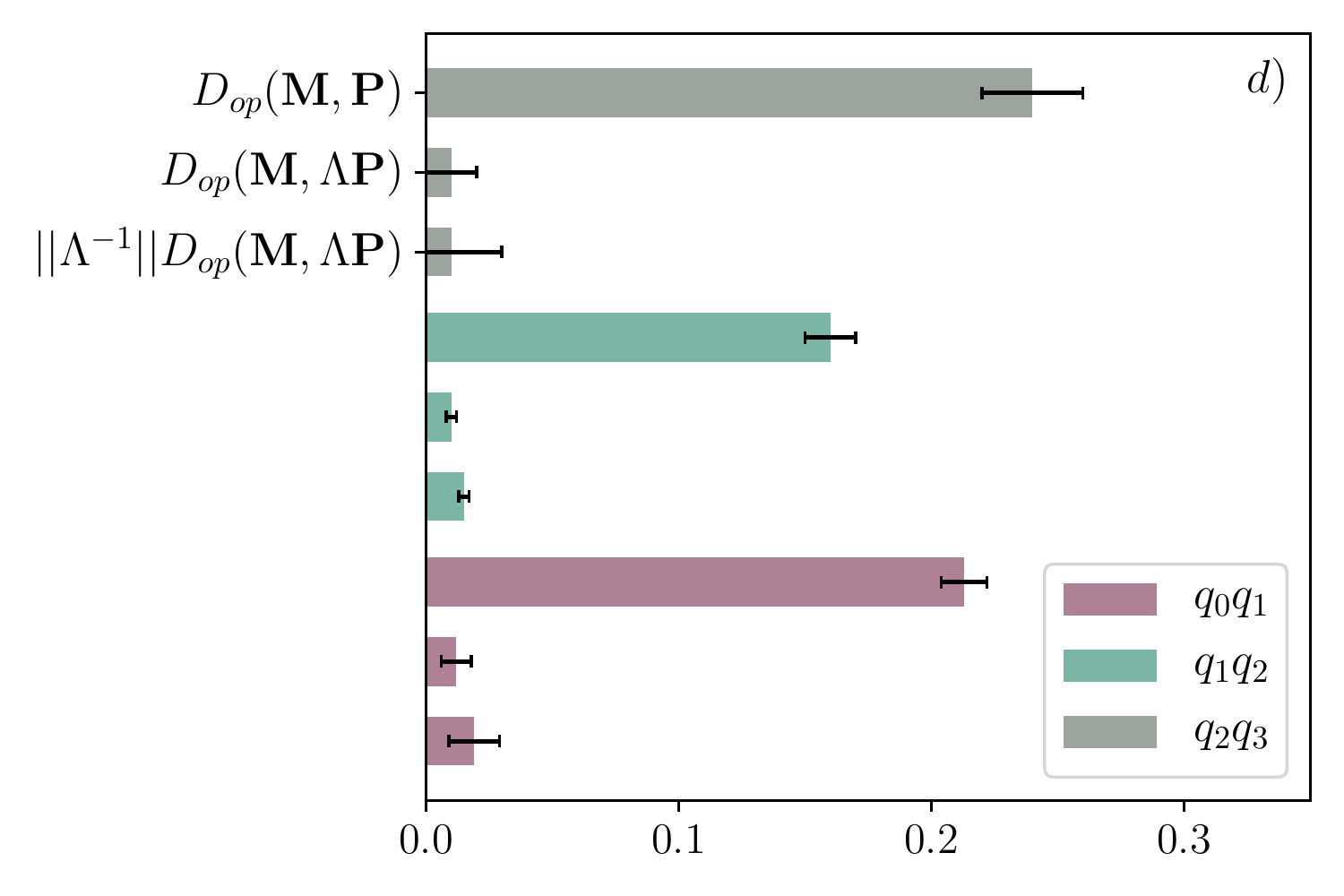}
		\label{fig::QDT_2q_b_rigetti}
		\caption{ \label{fig::QDT_2q}
			Barplot of the characterization of two-qubit measurements on a,b) all physically connected pairs of IBM's \textit{ibmqx4} and c,d) three exemplary physically connected pairs of Rigetti's \textit{Aspen-4-16Q-A}. 
			The presented data is analogous to Fig.~\ref{fig::QDT_1q}. On plots 
			a,c), the results were obtained via a joint two-qubit detector tomography, while on plots 
			b,d), the results were obtained via two single-qubit QDTs and the two-qubit POVMs created via tensor products.
			Each tomography experiment consisted of 36 circuits: all tensor products of Pauli eigenstates.
			Each circuit was implemented $8192$. Furthermore, each QDT experiment was repeated 4 times, in order to estimate the standard deviations, the corresponding $3\sigma$ bars are shown on the plot.
		}
	\end{figure*}
	
	\subsection{Quantum Detector Tomography}
	Here we present results of detector tomographies performed on all 5 qubits of IBM's five-qubit device \textit{ibmqx4} and on the first 5 qubits of Rigetti's 16-qubit device \textit{Aspen-4-16Q-A}.
	
	\subsubsection{Single qubit}

	In Fig.~\ref{fig::QDT_1q} we present the results of QDT performed on individual qubits in IBM's and Rigetti's devices. 
	Three observations can be made. 
	First, readout noise is significant for both platforms.
	We note that although it cannot be directly compared to the single-qubit gate errors from RB experiments, in the case of IBM  \textit{ibmqx4} average gate fidelities are so large that the readout-noise can be considered as the predominant type of noise on the level of individual qubits  (at least for short quantum circuits).
	Second, the distance between actual POVM and its diagonal form is in all cases small, which confirms the assertion that classical noise is the dominant type of noise in both devices. Lastly, we observe that the upper bound for our correction error $||\Lambda^{-1}||_{1\rightarrow 1} D_{op}\left(\M^{\text{exp}},\Lambda\M^{\text{ideal}}\right)$ (see Eq. \eqref{eq::error_diag})  is always significantly smaller than the distance of the noisy POVM from the ideal detector, which suggests that the mitigation scheme should be beneficial\footnote{We note that this upper bound can be used under the assumption of infinite statistics. We would like to remark that similar conclusions can be also made with regards to more conservative indicators of errors that take into account a limited number of performed experiments.}.

	We also visualize the results of QDT of single-qubit projective measurements on Fig~\ref{fig::1q_arrows}.  To this end , we parametrize the first effect of a measurement as $M_1=\sum_{k\in\lbrace{0,x,y,z\rbrace}} n_k\sigma_k$, and $\sigma_0\coloneqq \iden$. Then, we set $|z|\coloneqq \sqrt{n_x^2+n_y^2}$ to be the magnitude of coherent errors. In order for $M_1$ to remain element of POVM, its coefficients must satisfy  $n_0^2-1\leq |z|^2+n_z^2\leq n_0^2$.

	\subsubsection{Two qubits}
	We have also performed QDT for two-qubit projective measurements on Rigetti's and IBM's devices.
	In Figs.~\ref{fig::QDT_2q}\textcolor{red}{a, c} we present results of  detector characterization performed on the basis of joint two-qubit measurement tomography.  
	We also give results of the  alternative approach in which two qubit POVMs are reconstructed from tensor products of single-qubit POVMs 
	(see Figs.~\ref{fig::QDT_2q}\textcolor{red}{b, d})
	. 
	In both cases, the data resembles single-qubit scenario - coherent errors are relatively small and correction error $||\Lambda^{-1}||_{1\rightarrow 1} D_{op}\left(\M^{\text{exp}},\Lambda\M^{\text{ideal}}\right)$ (relevant for the infinite statistics scenario) is always much smaller than operational distance of the POVM from ideal detector.
	
	\begin{table}[]
		\begin{center}
			\includegraphics[scale=0.14]{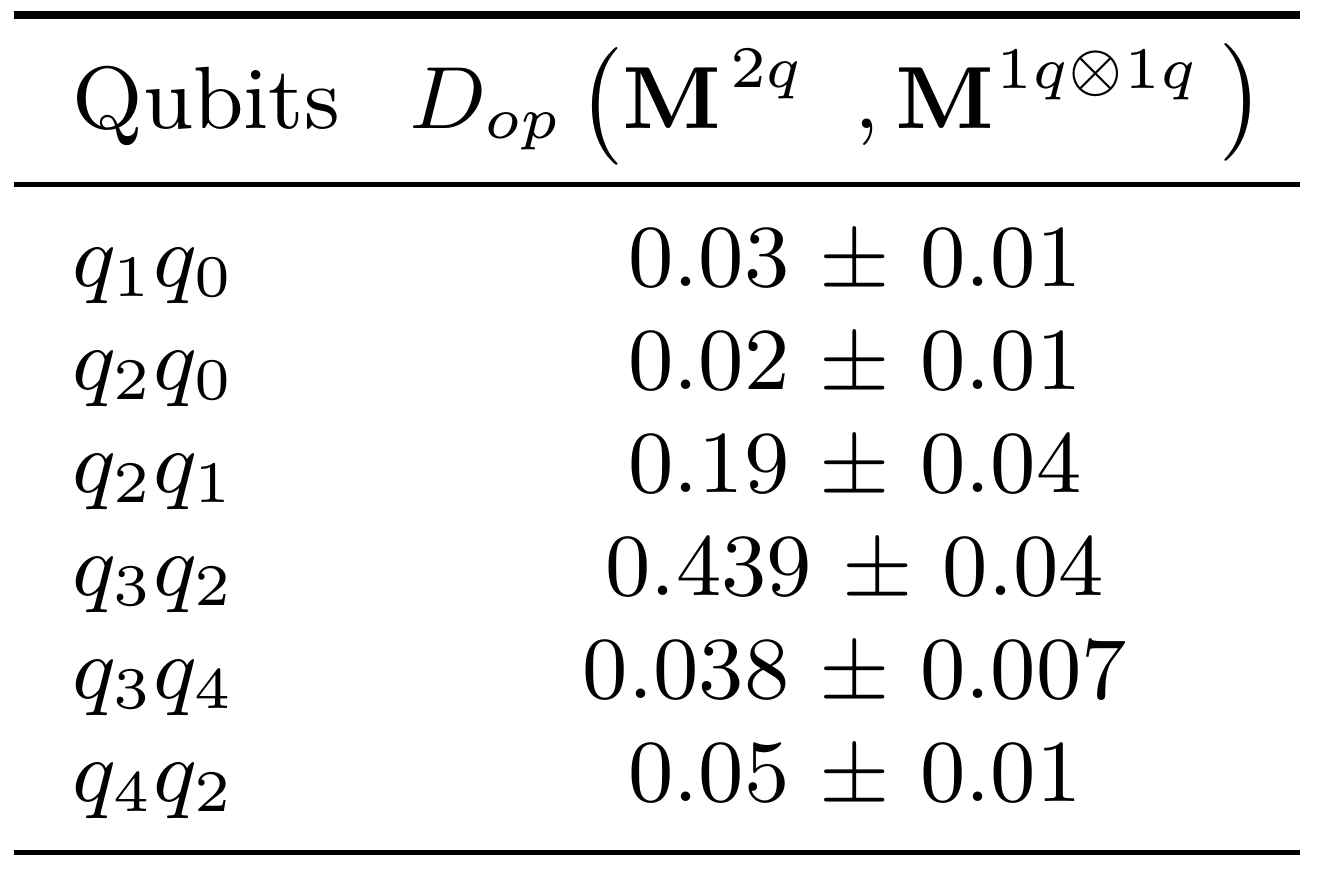}
			\includegraphics[scale=0.14]{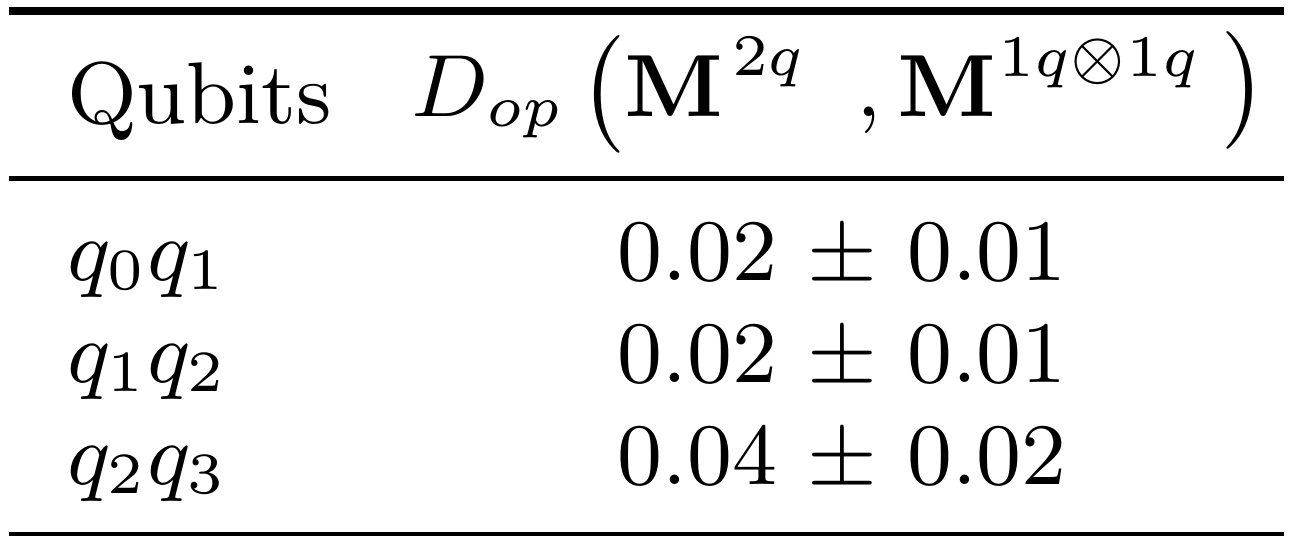}
		\end{center}
		\caption{Operational distances between the two-qubit POVMs that are obtained via the two-qubit QDT and the ones that are obtained via the tensor product of single-qubit POVMs. These values are provided for all physically connected qubit pairs in IBM's \textit{ibqmx4} (upper table) and for three exemplary physically connected pairs of Rigetti's \textit{Aspen-4-16Q-A} (lower table). 
			Each tomography was repeated 4 times in order to estimate standard deviations, the corresponding $3\sigma$ intervals are also presented in table.\label{tab::correlations}} 
	\end{table}
	We have also investigated the correlations between measurement errors. Tab.~\ref{tab::correlations} shows operational distances between POVMs reconstructed via two-qubit tomography and these obtained via tensor product of single-qubit measurements. We observe that for most pairs in IBM's device, the correlations between readout errors are small. However, with two exceptions of the pairs $q_3q_1$ and $q_2q_1$.
	This suggests that mitigation for those pairs should be performed based on joint two-qubit QDT. In the case of Rigetti's device, examined pairs do not show significant level of correlations.

	\section{Applications on IBM devices}\label{sec: Applications}
	
	We have performed numerous experiments on IBM \textit{ibqmx4} quantum device to demonstrate the usefulness of our error mitigation method for a number of quantum information tasks.
	In what follows we first describe briefly the theoretical basis of each of the tasks performed. Then, we proceed to the detailed description of the experimental results.
	\subsection{Theoretical description}
	\subsubsection{Quantum State Tomography \\and Quantum Process Tomography}
	Both Quantum State Tomography (QST) and Quantum Process Tomography (QPT) are based on the same idea as  Quantum Detector Tomography, with the clear difference that in the case of QST a quantum state is to be reconstructed, while a QPT provides the characterization of a quantum process, i.e., channel. 
	In our experiments, the tomographic reconstruction was done using the algorithm from \cite{Smolin2012}, which is available in qiskit. 
	
	To compare the target pure state $\ketbra{\psi}{\psi}$ with the tomographically reconstructed $\rho$ we used state fidelity \cite{mike&ike},
	\begin{align}\label{eq::state_fidelity}
	F^{state}(\ketbra{\psi}{\psi},\rho )=\Tr{\ketbra{\psi}{\psi}\rho}\ .
	\end{align}
	Likewise, in compare the target unitary channel $U$ to-be-ideally-implemented with the actual quantum channel $\Lambda$ obtained in the process tomography, we used the entanglement fidelity \cite{Schumacher1996,Horodecki1998}, which may be calculated as
	\begin{align}
	F^{ent}=\Tr{\Phi_{\text{U}}\rho_\Lambda},
	\end{align}
	where $\Phi_{\text{U}}$ is the Choi-matrix representation of the unitary channel and $\rho_\Lambda$ is the Choi-matrix representation of $\Lambda$.

	\subsubsection{Implementation of non-projective measurements}
	In our work, we have used generalized quantum measurements as models for noisy projective measurements.
	However, it is certainly not the only application of POVMs in quantum information.
	Indeed, POVMs can outperform projective measurements in some tasks, such as unambiguous state discrimination \cite{Chefles1998, Barnett2009}, minimal error state discrimination \cite{Chefles1998, Barnett2009}, quantum tomography \cite{Renes2004}, port-based teleportation of quantum states \cite{Ishizaka2008} or quantum computing \cite{RevModPhysComp2010}. 
	Therefore, it may happen that one actually wants to \textit{implement} a certain generalized measurement. 
	
	The standard method to do so is via the Naimark's extension \cite{Peres2006,Preskill2012} and requires adding an ancilla system and performing a projective measurement on a whole system. 
	We have implemented in this way three different single-qubit POVMs. 
	We have assessed the quality of implementation by performing quantum measurement 
	tomography and computing the operational distance, see Eq.~\eqref{eq::op_distance}.
	
	\subsubsection{Quantum algorithms -- Grover's search and the Bernstein-Vazirani algorithm}
	Grover's \cite{Grover1996} and Bernstein-Vazirani \cite{Bernstein1993} algorithms are the canonical examples of quantum algorithms operating withing an oracular (black box) model of computation. 
	Grover's search aims to find  unknown element $y$ encoded in the application of the unitary gate $U_y$ defined via $U_y \ket{x} = (1-2\delta_{x,y})\ket{x}$, where $\delta_{x,y}$ is the Kronecker's delta.
	BV algorithm algorithm uses Hadamard transform to find (in a single query) a hidden string $s\in\mathbb{Z}_{2}^n$ encoded in a  $n$-qubit  unitary transformation $V_s$ defined via  $V_s\ket{x}=(-1)^{s.x}\ket{x}$, where $s.x=\sum_{i=1}^n s_i x_i\ (\text{mod }2)$. 
	
	We have based our implementation of these routines on the expository work \cite{Coles2018}. Both algorithms were implemented on three qubits, one of which was an ancilla. In that case, Grover's algorithm required only a single query to the oracle, hence it could have been realized in a single quantum circuit. The figure of merit for the quality of implementation in the case of both algorithms is a single number -- the probability of obtaining a particular outcome. We have used this number to benchmark our error mitigation scheme.

	\subsubsection{Probability distributions}
	We have implemented various five qubit circuits in the IBM device aiming to generate specific probability distributions upon measuring all the qubits.  As a figure of merit we have used Total Variation distance (see \eq{eq::tv_distance}) between target probability distribution, and the one estimated from relative frequencies. In order to test our error mitigation method, we have implemented these probability distributions with and without the error-mitigation procedure.
	
	\subsection{Experimental results}
	Now we present experimental results that demonstrate the practical effectiveness of our error-mitigation procedure. 	
	For each presented experiment, QDT from which we inferred correction matrix $\Lambda^{-1}$, has been performed in the same calibration period as the corrected experiments.
	Moreover, for tasks involving measuring multiple-qubits, we compare mitigation which assumes a lack of correlation between qubit readout errors, with the one which accounts for such correlations.

	\subsubsection{Quantum State Tomography \\and Quantum Process Tomography}\label{subsec:QST_QPT}

	In what follows we present results of the quantum state and quantum process tomographies performed on single-qubit systems and quantum state tomographies performed on two-qubit systems. 
	The procedure of tomography involves performing multiple experiments that are then used to reconstruct the objects in question. 
	We use an error-mitigation procedure to correct the statistics (probability vectors) in these experiments in order to enhance the quality of the tomographic reconstruction.
	
\begin{figure}[h!]
		
		\begin{subfigure}
			\centering
			\includegraphics[scale=0.55]{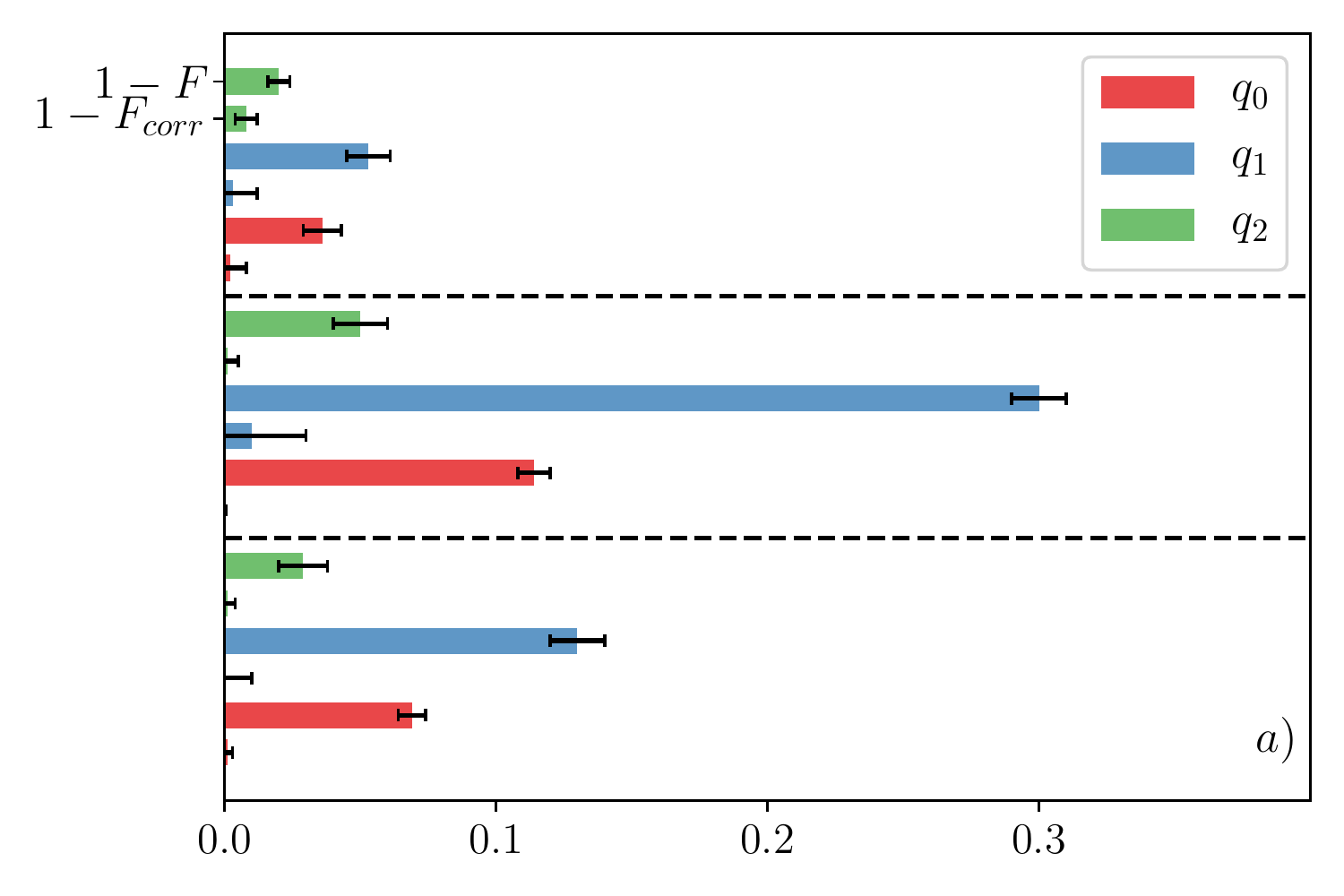}
			\label{fig::QST_1q}
		\end{subfigure}
		\hspace{0.005cm}
		
		\begin{subfigure}
			\centering
			\includegraphics[scale=0.55]{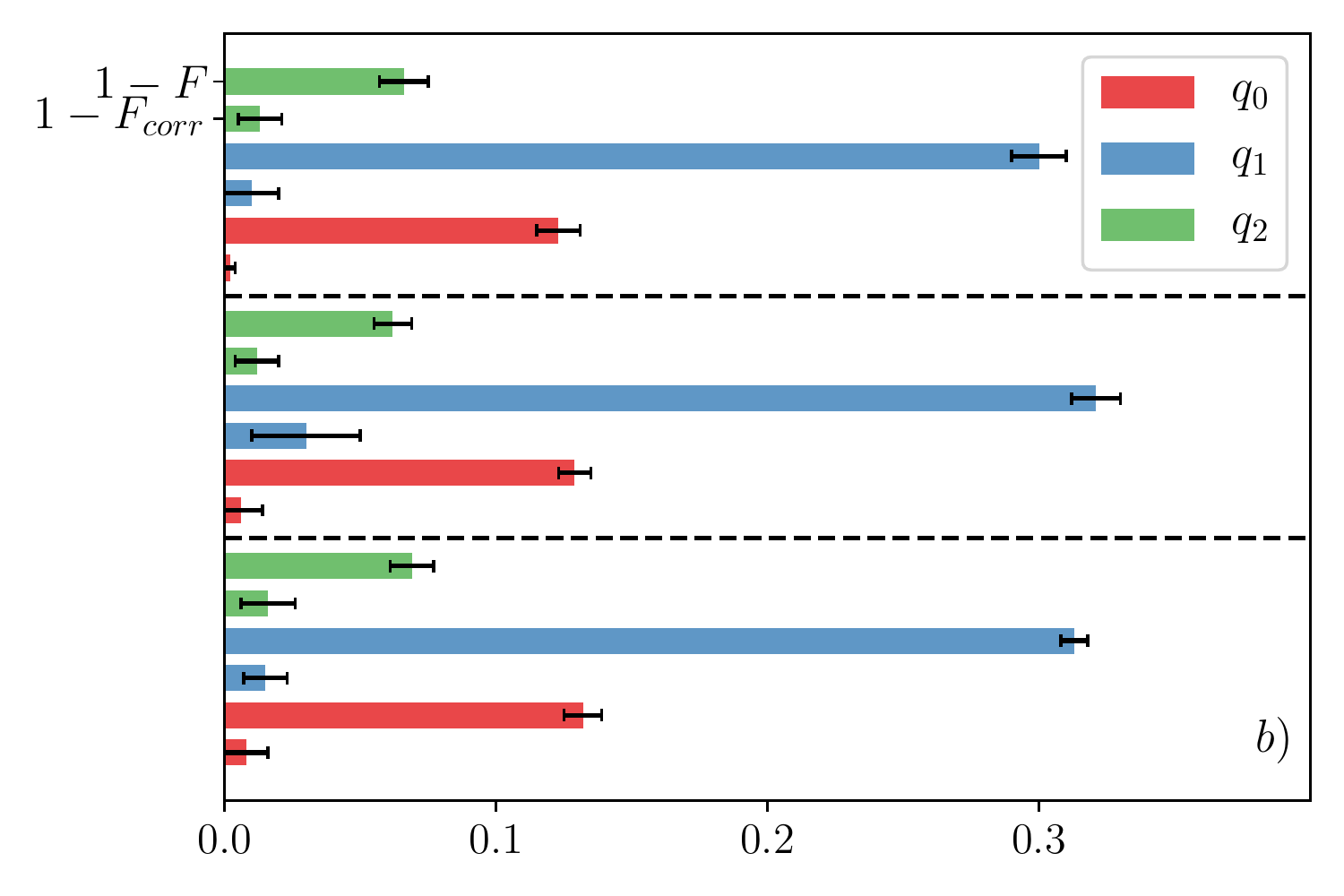}
			\label{fig::QPT_1q}
		\end{subfigure}
		\hspace{0.001cm}
		\caption{ \label{fig::QST_QPT_1q}
			Infidelities obtained from the reconstruction of a) single-qubit states and b) single-qubit channels.
			Different colors refer to different qubits.
			The first bar in each pair corresponds to uncorrected statistics, while the second bar corresponds to the corrected one.
			The three regions separated by dashed lines refer to different Haar-random \cite{online_repo} quantum states (processes) to be reconstructed.		
			Each tomography experiment consisted of several quantum circuits (3 for QST, 12 for QPT), and each quantum circuit was implemented $8192$ times. 
			Furthermore, each tomography experiment was repeated 5 times in order to estimate statistical errors, the corresponding $3\sigma$ intervals are shown on the barplot as black lines. The correction  was based on a QDT which used an over-complete set of states (all Pauli eigenstates), with each circuit implemented $32768$ times.}
	\end{figure}

	In Fig.~\ref{fig::QST_QPT_1q} we present results for different quantum state and process tomographies on individual qubits in IBM's \textit{ibmqx4} device. 
	It is worth noting that although results highly depend on the input state, our correction reduces infidelities of reconstructed states in every tested case.
	
\begin{figure}[t!]
		%		\begin{subfigure}
		%			\centering
		%			\includegraphics[scale=0.52]{fig_QST_1q1q.pdf}
		%			\label{fig::QST_2q_a}
		%		\end{subfigure}
		%		\begin{subfigure}
		%			\centering
		%			\includegraphics[scale=0.52]{fig_QST_2q.pdf}
		%			\label{fig::QST_1q1q}
		%		\end{subfigure}
		\includegraphics[scale=0.54]{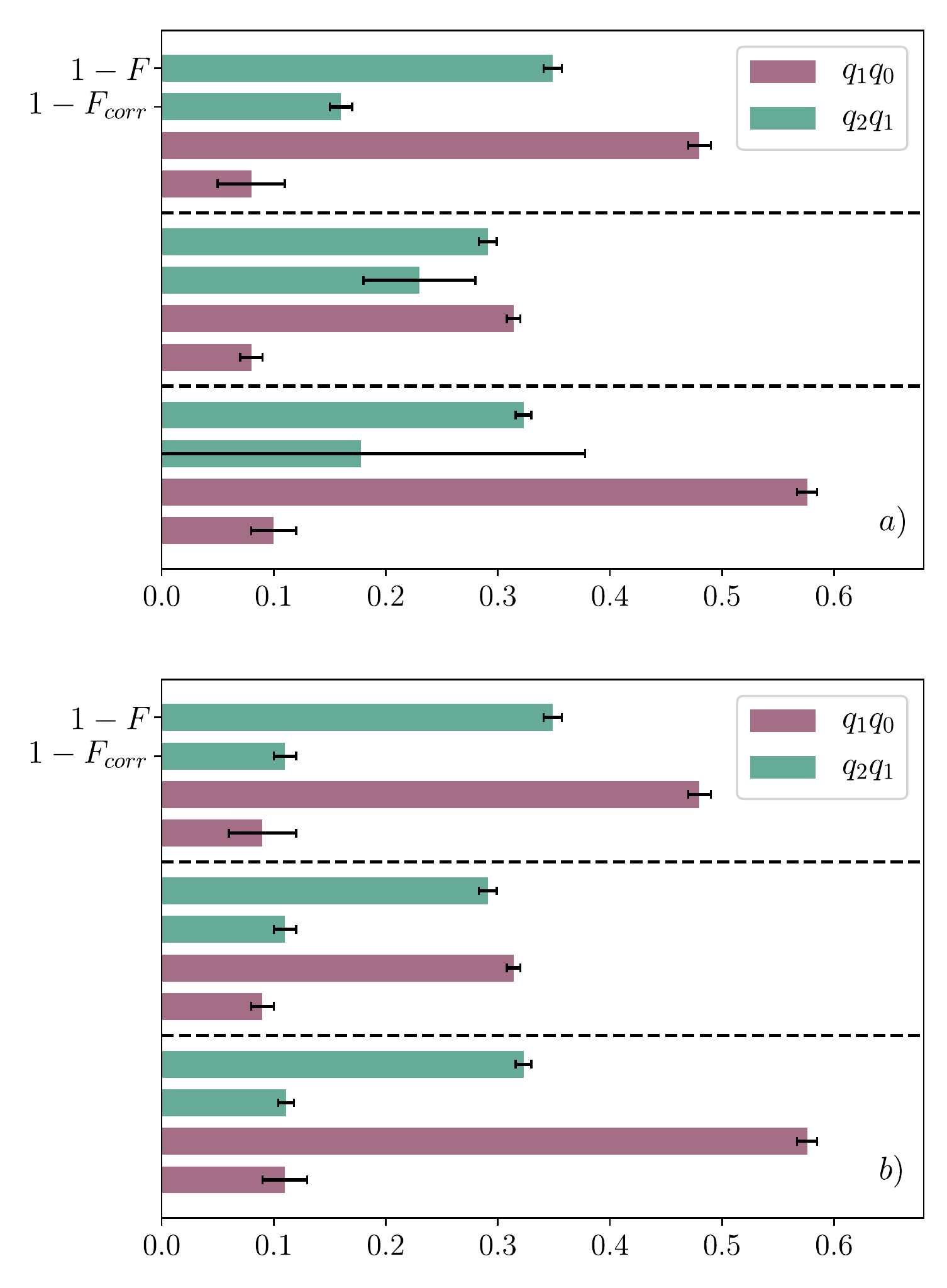}
		\caption{ \label{fig::QST_2q}
			Infidelities obtained from the tomographic reconstruction of Haar-random \cite{online_repo} two-qubit quantum states using error-mitigation procedures based on a) uncorrelated single-qubit QDTs and b) correlated two-qubit QDT. 
			The convention for presenting data is analogous to that of Fig.~\ref{fig::QST_QPT_1q}. 
			Each tomography experiment consisted of 9 quantum circuits, corresponding to local measurements performed in different Pauli bases. Each quantum circuit was implemented $8192$ times. Furthermore, each tomography experiment was repeated 5 times in order to estimate statistical errors,  the corresponding $3\sigma$ intervals are shown on the barplot as black lines. The QDT used in the error-mitigation procedure was implemented using an over-complete set of states (all Pauli eigenstates), with each circuit implemented $32768$ times.
		}
	\end{figure}

	We observe the improvement also for two-qubit state tomographies, as shown in Fig.~\ref{fig::QST_2q}. 
	We compare the performance of the error-mitigation procedures based on the measurement reconstruction obtained from two-qubit QDTs and the one obtained from the tensor product of single-qubit QDTs. 
	Interestingly, although one would expect that for highly correlated readout errors (in particular the pair $q_2q_1$), accounting for those correlations in readout errors should provide an improvement over the mitigation based on not-correlated tomography, we observed such improvement in the case of only one out of three tested quantum states.
	
	We remark on the possible causes of the result reported above. 
	First, in the course of correction of statistics, we sometimes obtain nonphysical probability vectors.
	As already pointed out in Section \ref{sec: Errors_c} dealing with such situations potentially introduce additional errors. 
	These errors can be different for mitigation schemes based on different tomographic reconstructions. 
	The second probable cause is the insufficient number of experiments necessary to perform a reliable QDT of a two-qubit measurement (the sample complexity of this problem is definitely higher as there are much more parameters needed to describe the two-qubit measurement compared to two single-qubit measurements).

	\subsubsection{Implementation of non-projective measurements}
	\begin{figure}[h!]
		%		\begin{subfigure}
		%			\centering
		%			\includegraphics[scale=0.52]{fig_naimark_1q1q.pdf}
		%			\label{fig::naimark_a}
		%		\end{subfigure}
		%		\begin{subfigure}
		%			\centering
		%			\includegraphics[scale=0.52]{fig_naimark_2q.pdf}
		%			\label{fig::naimark_b}
		%		\end{subfigure}
		%	
		\centering
		\includegraphics[scale=0.52]{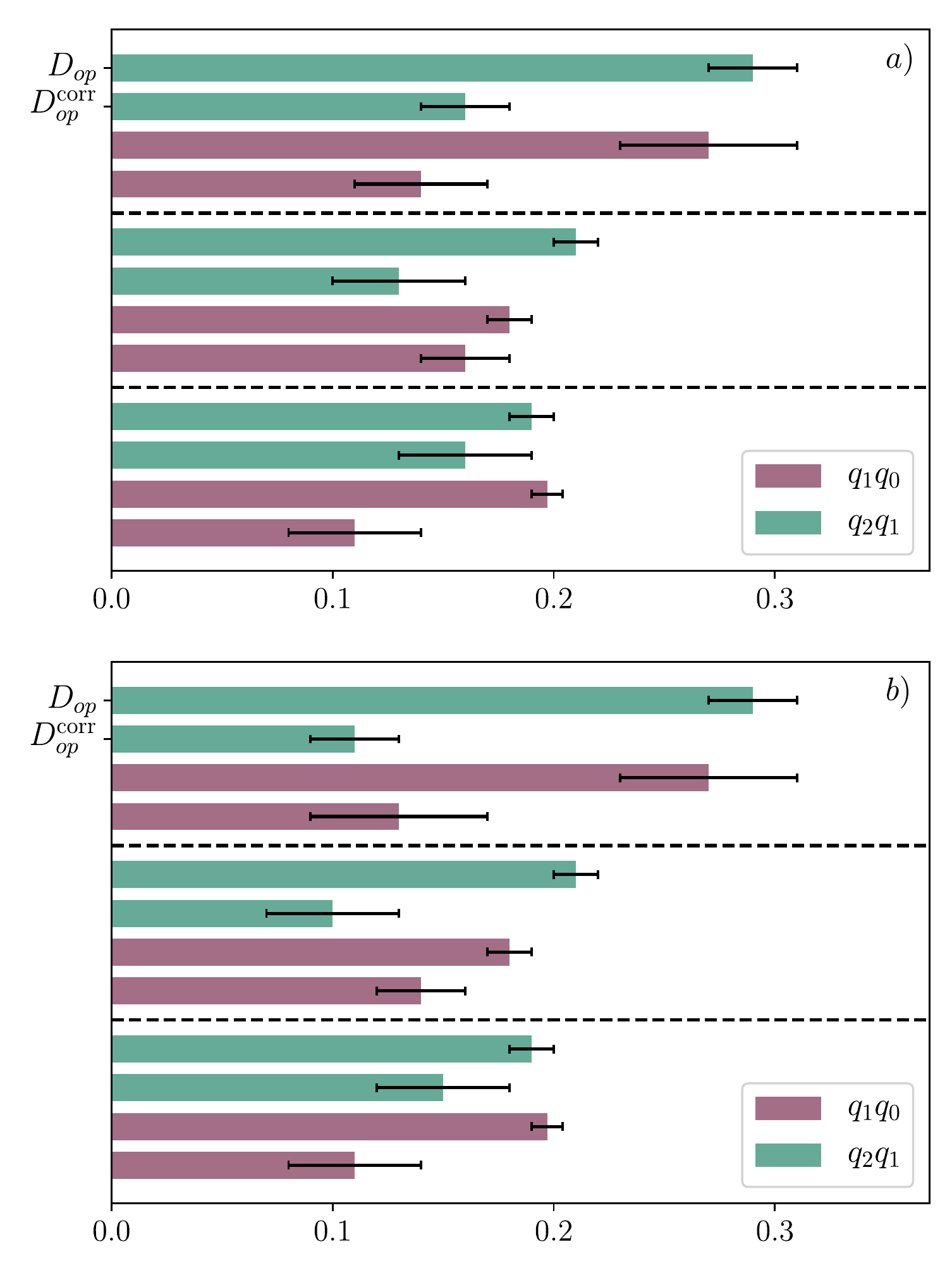}
		
		\caption{Operational distances for POVMs implemented in \textit{ibmqx4} via Naimark's extension. 	
			Different colors correspond to different pairs of qubits.
			The first bar in each pair depicts the operational distance between the to-be-implemented POVM and the one obtained in the measurement tomography based on uncorrected statistics ($D_{op}$); while the second bar shows the distance between the ideal POVM and the one reconstructed from a tomography based on statistics corrected by our scheme ($D_{op}^{corr}$).
			The three regions separated by dashed lines correspond to different to-be-implemented POVMs: the Haar-random 4-outcome POVM, the tetrahedral measurement, and the trine measurement \cite{online_repo}.
			Each POVM reconstruction required 4 quantum circuits (Pauli 'x+','y+','z+' and 'z-' states) and each quantum circuit was implemented 8192 times. 
			Reconstructions were repeated 5 times in order to estimate standard deviations, the corresponding $3\sigma$ intervals are shown here as black bars.
			The mitigation or measurement errors by our method was done based on a) separate single-qubit QDTs, and b) joint two-qubit QDT.
			The QDTs were based on the preparation of Pauli 'x+','y+','z+' and 'z-' states for single qubits, and all of their tensor-product combinations for pairs of qubits. 
			Each circuit for the QDT was implemented $32768$ times.}\label{fig::naimark}
	\end{figure}
	
	For the implementation of non-projective measurements via Naimark's extension, we observe the general improvement of the quality of reconstructed POVMs when the error-mitigation scheme is used (see Fig.~\ref{fig::naimark}). Moreover, the difference between error-mitigation procedure based on the non-correlated and on correlated detector tomography is negligible this time. 
	However, we can offer a simpler explanation than in the case of two-qubit QSTs described above. 	
	Namely, at the time when those particular experiments were performed, correlations between errors for pair $q_2q_1$ were relatively small (compared to data regarding quantum process tomography and quantum detector tomography). 
	This points to the importance of performing systematic device characterization and calibration.

	\subsubsection{Quantum algorithms -- Grover's search and the Bernstein-Vazirani algorithm}

	\begin{table}[h!]
		\begin{center}
			\includegraphics[scale=0.18]{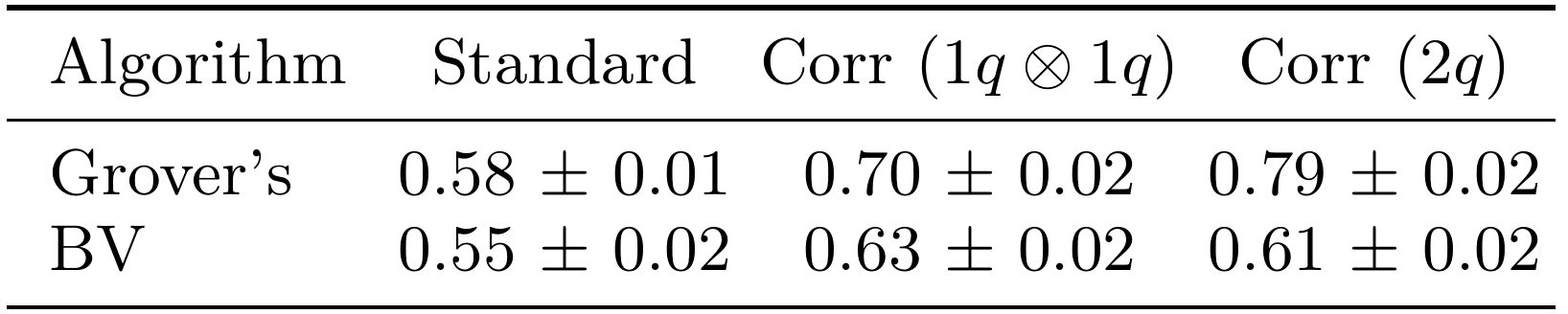}
		\end{center}
		\caption{The guess probabilities for the three-qubit Grover's search and BV algorithms implemented on \textit{ibmqx4}. 
			The third column shows the mitigation based on uncorrelated single-qubit QDTs, and the fourth column shows mitigation where the correlations between readout errors were accounted for.
			For both algorithms, we used the qubits $q_0$, $q_1$, and $q_2$. 
			The measurement was done on the  $q_2q_1$ pair for Grover's search and on the pair $q_1q_0$ for the BV algorithm. For Grover's search, the hidden element was chosen to be  '11', while for BV it was '01'. 
			Each experiment was repeated 5 times in order to estimate standard deviations, the corresponding $3\sigma$ intervals are given in the table.
			The mitigation was based on a minimal QDT (Pauli 'x+','y+','z+' and 'z-' states), with each circuit implemented $32768$ times.}\label{tab::algorithms}
	\end{table}

	\begin{table}[h!]\centering
		\includegraphics[scale=0.21]{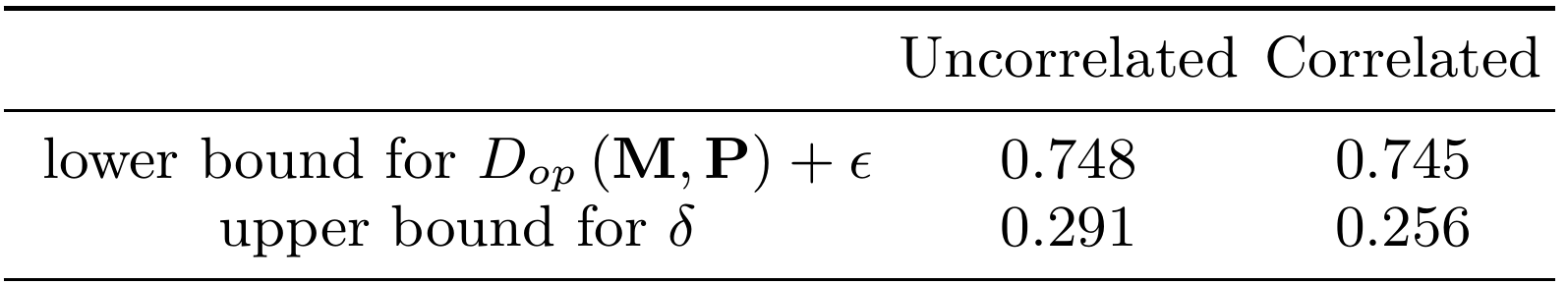}
		\caption{Errors for five-qubit measurements in \textit{ibmqx4}. 
			The first row presents the \emph{lower bound}\footnote{Calculating operational distance (Eq.~\eqref{eq::op_distance}) via search over all possible combinations of indices quickly becomes computationally infeasible. However, a lower bound can be found by exhaustive search over a sufficiently large subset of all combinations. On the other hand, assuming a tensor structure of POVMs in consideration, an upper bound can be easily obtained using the subadditivity of the operational distance (see Appendix~\ref{app::proofs} for details). Finally, note that to estimate whether the inequality in Eq.~\eqref{eq:RULEcorr} is satisfied, indeed the upper bound for LHS and the lower bound for RHS is needed.} for the operational distances between noisy and ideal measurements \emph{including} statistical errors $\epsilon$ with the number of repetitions  $N= 72 \times 8192=589824$ and accepted error probability $\mathrm{Pr}_{err}=0.01$ (see Eq.~\eqref{eq::statistical_error}). The second row gives the \emph{upper bound} on $\delta$ defined in Eq.~\eqref{eq::error_diag_stat}, which bounds the errors in the observed statistics.
			The second column corresponds to a tomographic reconstruction based on the tensor product of five single-qubit QDTs, while in the third column the joint QDT for a pair $q_2q_1$ was included. 	
			The presented data is for POVMs which were used for the correction procedure, hence we do not provide standard deviations.
		}\label{tab::5q_dops}
	\end{table}	
	
	From the data presented in Table \ref{tab::algorithms} we observe that our error mitigation method is favorable for the implementation of both Grover and BV algorithms.
	In the case of correction performed on the highly correlated pair $q_2q_1$ (Grover's search), the mitigation scheme accounting for the correlations of readout errors provided a significant advantage over the scheme using non-correlated tomography. 
	
	We remark that that relative improvement is much higher in the case of Grover's algorithm than in BV's. 	
	This may result from two factors.
	First, the circuit for Grover's algorithm has lower depth than the circuit for the BV algorithm, therefore, in this case, gate-errors might have a smaller impact on the final result. 
	Second, for the implemented instance of Grover's algorithm, the expected result was '11' (which is the state most prone to the noise observed in IBM's device), while for the BV algorithm the theoretically correct result was '01'.

	\begin{rem}
		We note that in Ref.~\cite{Coles2018}, authors obtained values $\approx 0.65$ and $\approx0.386$ for the correct guessing probabilities in the same instances of  Grover's and BV algorithms.
		Both of these values differ significantly from the ones observed by us without implementing the correction procedure. 
		Such discrepancy is another confirmation that indeed device's performance varies over long periods of time. 	
	\end{rem}

	\subsubsection{Probability distributions}
	
	Finally, we present results on the implementation of probability distributions on five qubits. We start by giving the operational distances and bounds on the correction errors for the five-qubit detector in Table~\ref{tab::5q_dops}. From the data presented there, it is clear that even for 5 qubits we are still in the regime in which correction is beneficial (see the criterion given in  Eq.~\eqref{eq:RULEcorr}).
	\begin{table}[t!]
		\centering
		\includegraphics[scale=0.18]{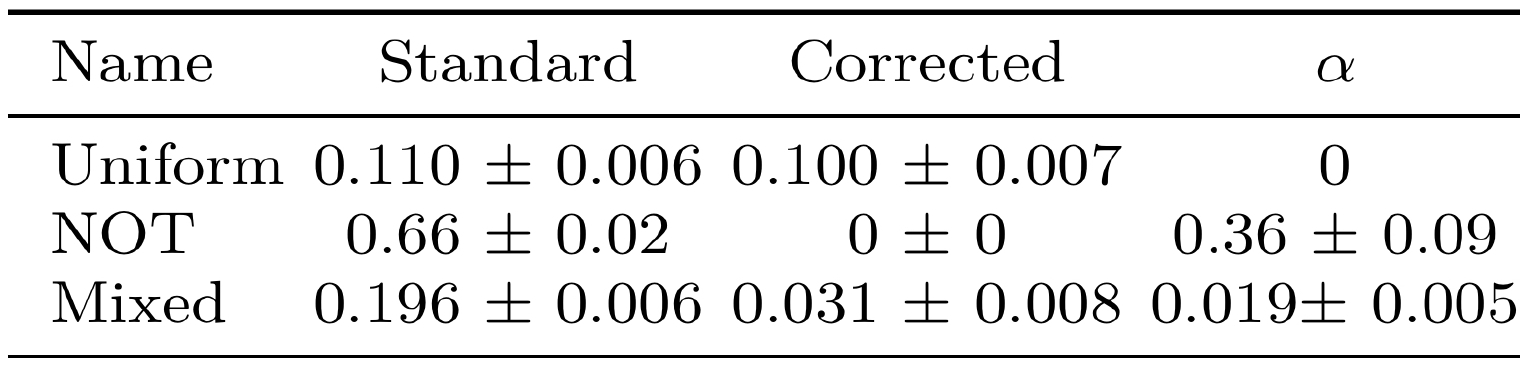}	\label{PD_Fig.s_5q_a}
		\vspace{2mm}
		\includegraphics[scale=0.18]{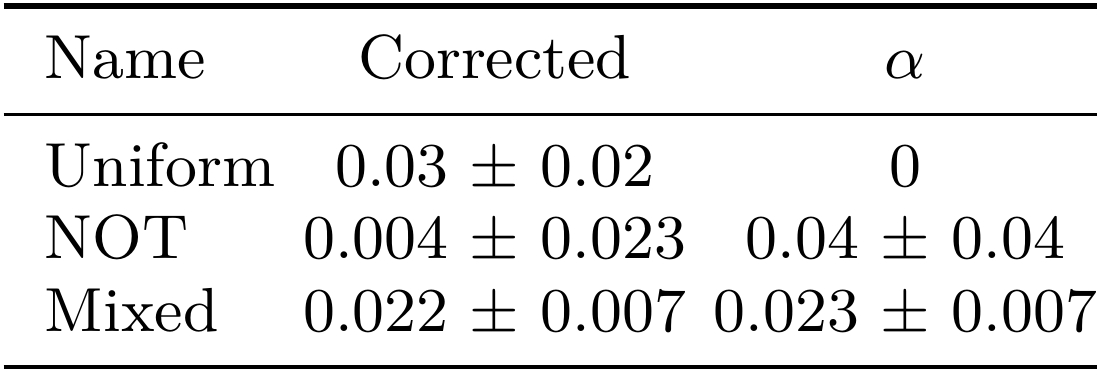} \label{PD_Fig.s_5q_b}
		\caption{\label{tab::PD_5q} Results of experiments for implementing specific probability distributions on the \textit{ibmqx4} processor. We use the total variation distance between the target and the reconstructed probability distribution in order to asses the quality of the implementation. The upper table presents results of for the uncorrected case and the case when applying a correction procedure that does not take into account the correlations in the readout errors, while the lower table provides the results of a correction procedure where the correlations between a single qubit pair ($q_2q_1$) measurements were taken into account.
			Each experiment consisted of a single quantum circuit, which was run $72*8192=589824$ times. 
			This high number of repetitions was necessary in order to minimize the statistical errors.
			Furthermore, each experiment was repeated four times in order to estimate the statistical errors. The errors given in the tables have magnitude $3\sigma$, where $\sigma$ is the estimated standard deviation.  
			The correction procedure used in the experiments was based on over-complete (all Pauli eigenstates) single-qubit QDTs with each circuit implemented $32768$ times. 
			The two-qubit QDT for pair $q_2q_1$ was done using all tensor-product combinations of Pauli eigenstates on both qubits, with each circuit implemented the same number of times. }
	\end{table}

	The probability distributions tested here were generated by performing a simple circuit (acting on the standard input state $\ket{0}^{\otimes 5}$ ) followed by a measurement in the computational basis. 	 We have chosen to implement the following three probability distributions:
	The first distribution was the uniform one and was implemented by the tensor product of five  Hadamard gates.  The second probability distribution, named  '\textit{NOT}', was implemented by the product of five X gates. 	For the last probability distribution, named by us '\textit{Mixed}', the circuit consisted of two X gates on $q_0$ and $q_2$, two Hadamards on $q_3$ and $q_4$, and the identity gate on $q_1$.
	
	Results of the correction of the probability distributions are presented in Tab.~\ref{tab::PD_5q}. There are a few interesting features to be observed. First, let us note that in the case of the uniform probability distribution, our correction procedure is ineffective if we do not account for correlations between qubit readout errors. However, if we do account for such correlations in only a \emph{single} pair of qubits (the highly correlated pair $q_2q_1$), it is clear that correction procedure provides a high improvement\footnote{
		It is important to note, that although from Table ~\ref{tab::correlations} one can see that there is also another highly correlated pair, $q_3q_2$, our experiments for the five-qubit case were performed at the time in which only the pair $q_2q_1$ has shown a significant level of correlations.
		See Appendix~\ref{app::exp_data} for the dates of execution of all experiments.}.	
	Second, in the case of the 'NOT' probability distribution, the correlation-ignorant correction has attained the perfect estimation -- yet with remarkably high distance $\alpha$ between first-guess non-physical probability vector and its closest physical neighbor. This moves us very close to the regime in which the correction is unreliable according to the criterion given in Eq.~\eqref{eq:RULEcorr}. 
	Accounting for correlations this time proves to improve results also remarkably highly, this time with only slightly distant first-guess non-physical distribution. Finally, the method provides an advantage for the 'Mixed' distribution in both cases. 
	However, interestingly, this time the difference of improvement after accounting for correlations is negligible.

\section{Conclusions and further research directions} \label{sec: Conclusions}
\subsection{Conclusions}

In this work, we have presented a scheme for the mitigation of readout errors which is suitable for noisy and imperfect quantum devices. 
Our method uses solely classical post-processing that depends on the results obtained in the course of  Quantum Detector Tomography. In principle, it can be applied on any quantum hardware provided the readout errors are classical and one has access to approximately perfect detector tomography. 
We have analyzed how our error-mitigation procedure is affected by non-classical noise and finite statistics (at the stage of estimation of probability distributions). 
We have also presented a comprehensive study confirming the (approximate) validity of the classicality of errors in the publicly available prototypes of quantum chips based on architectures with superconducting transmon qubits. 

Our method was tested on a variety of quantum tasks and algorithms implemented on IBM's five-qubit quantum processor. 
We observed that our scheme, despite relying on purely classical processing of the obtained results, yields substantial improvements in the performance for a number of in a single-, two- and five-qubit experiments on the \textit{ibmqx4} device.
We have compared the error mitigation scheme which accounts for correlations in readout errors with the one that does not account for them. 
Taking into account correlations proved to be crucial in the case of experiments with single probability vectors (Grover's algorithm for two qubits and the probability distributions in the five-qubit case). 
We point at the study of readout error correlations and accounting for them as a future research direction.

We are convinced that our method and its future extensions will find applications to mitigate readout errors in realistic near-term quantum devices that, despite being inherently noisy and imperfect, are expected to be capable of effectively solving problems of practical relevance. One can argue that performing quantum detector tomography and implementing our error-mitigation procedure will not be feasible for systems involving many qubits (due to the exponential growth of the size of this problem). 
However, many quantum algorithms tailored to near-term applications (such as quantum approximate optimization algorithm \cite{QAOA2014} or variational quantum eigensolvers \cite{VQE2017IBM}) can be performed by implementing just a polynomial number of few-qubit quantum measurements. We expect that our scheme can be particularly useful in such scenarios.

\subsection{Further research}
Finally, we state a number of possible future research directions.

The first question concerns the possibility of \emph{physical} correction of coherent readout errors. 
For the case of a single-qubit measurement there exists a natural possibility of correction of the coherent measurement errors.
Namely, if the readout errors are uncorrelated one could obtain single-qubit POVMs with diagonal effects by implementing suitably chosen unitaries at the end of every quantum circuit. 
Moreover, in the case of uncorrelated errors, such physical correction is in principle possible for an arbitrary number of qubits (provided that readout errors dominate over gate errors). Determining whether this procedure will work for realistic near-term devices is an important research question.

Another important problem is to give confidence intervals (with respect to the operational distance defined in Eq. \eqref{eq::DopDtv}) for the problem of quantum detector tomography.
This kind of results is required to give estimates for the sample complexity of QDT and hence are crucial in assessing the feasibility of QDT on realistic near-term quantum devices. 
So far, such results have been obtained only for the case of quantum state tomography and the trace distance as the figure of merit \cite{Guta2018,Wang2019}.

Last but not least, it is desirable to extend our error-mitigation procedure to larger systems with complicated geometry of the physical connections between qubits.  
To realize this goal it is necessary to efficiently perform detector tomography of multi-qubit devices and understand if the readout correction is possible without the necessity to perform estimation of the full probability distribution. 

\begin{center}
	\textbf{Code availability}
\end{center}
We are making Python code implementing the ideas presented in this work publicly available in the form of GitHub repository: \url{https://github.com/fbm2718/QREM}. 
We intend to further develop the repository in the future, adding new functionalities and tutorial Jupiter notebooks, which we hope will make our method more useful in practice.

\begin{acknowledgments}
	We thank \'A. Budai, R. Demkowicz-Dobrza\'nski, M. Horodecki, H. Pashayan, A. P\'alyi, and J. Tworzyd\l{}o  for interesting, fruitful discussions.
	We are especially grateful to \L{}. Pawela for running our code on the Riggetti device. 
	M.O.  acknowledges the support of Homing programme of the Foundation for Polish Science. F.B.M acknowledges the support by the Foundation for Polish Science through IRAP project (contract no.2018/MAB/5). Homing and IRAP programmes are co-financed by EU within
	Smart Growth Operational Programme.  Z.Z. was supported by the Hungarian National Research, Development and Innovation Office (NKFIH) through Grants No.  K124351, K124152, K124176 KH129601, and the Hungarian Quantum Technology National Excellence Program (Project No. 2017-1.2.1-NKP-2017- 00001); and he was also partially funded by the J\'anos Bolyai and the Bolyai+ Scholarships. 
	We acknowledge the use of the IBM Quantum Experience and Rigetti Forest SDK for this work. 
	The views expressed are those of the authors and do not reflect the official policy or position of IBM or the IBM Quantum Experience team, or the Rigetti team.
	
\end{acknowledgments}

\begin{center}
	\textbf{NOTE ADDED}
\end{center}
Upon completion of this manuscript, we became aware of a recent work \cite{Chen2019} that proposed an analogous scheme for the mitigation of readout errors. 
Furthermore, we note that IBM's qiskit package has been recently updated to include the readout error correction scheme, which seems to rely on an analogous procedure.

\printbibliography
%\onecolumngrid
%\appendix
\onecolumngrid
\appendix

\section{Proofs of technical statements}\label{app::proofs}
Here we prove technical results that were given without proofs in the main text. 
First, we present the proof of equality \eqref{eq:uncorrERR}, which we recall here for the Reader's convenience
\begin{equation}\label{eq:uncorrERR2}
D_{op}\rbracket{\M^{(K)},\P^{(K)}}=1-\prod_{i=1}^K\left(1-D_{op}\rbracket{\Lambda_i \P_i,\P_i}\right)\ .
\end{equation} 
In the above formula $\M^{(K)}=(\otimes_{i=1}^NK \Lambda_i ) \P^{(K)}$ is the noisy version of the multi-qubit projective measurement $\P^{(K)}=\otimes_{i=1}^K \P_i $ (see Section \ref{sec: Main} for more details). 
Directly from the definition of  the operational distance given in Eq.~\eqref{eq::DopDtv}, one obtains
\begin{equation}\label{eq:int1}
D_{op}\rbracket{\M^{(K)},\P^{(K)}}= \max_{\rho} D_{TV}\rbracket{\vec{p}^{\M^{(K)}},\vec{p}^{\P^{(K)}}}\ ,
\end{equation}
where $\vec{p}^{\M}$ is the probability vector of probabilities generated by the statistics of measurements of $\M$ on a quantum state $\rho$. 
Using the linearity of the Born rule in quantum states and the fact that both $\M^{(K)}$ and $\P^{(K)}$ are diagonal in the computational basis in $\left(\mathbb{C}^2\right)^{\otimes K}$, one can immediately prove that the optimal state in \eqref{eq:int1} can be also chosen to be \emph{diagonal} in the computational basis.
Moreover, from the linearity of the Born rule it also follows that $D_{TV}\rbracket{\vec{p}^{\M^{(K)}},\vec{p}^{\P^{(K)}}}$ is a convex function of $\rho$ and therefore the maximum in \eqref{eq:int1} is always attained for \emph{pure states}. 
Combining these two observations, we see that the optimal input state $\rho^\ast$ will be a particular computational basis state: $\rho^\ast=\ketbra{\underline{i}}{\underline{i}} =\ketbra{i_1}{i_1}\otimes \ketbra{i_2}{i_2} \otimes \ldots\otimes \ketbra{i_K}{i_K}$. 
The optimization over such states becomes now easy since the projective measurement $\P^{(K)}$ acts as the identity transformation on probability distributions (i.e., on diagonal quantum states). 
Therefore, we get 
\begin{equation}\label{eq:subadditivity}
D_{op}\rbracket{\M^{(K)},\P^{(K)}} = \max_{\underline{i}}  D_{TV}\left(\otimes_{i=1}^K \Lambda_i \ketbra{\underline{i}}{\underline{i}},\  \ketbra{\underline{i}}{\underline{i}} \right) =   1- \prod_{i=1}^K (1-\max\lbrace{p_i,q_i\rbrace})  \ ,   
\end{equation}
which, by the virtue of \eqref{eq:distance single} concludes the proof of \eqref{eq:uncorrERR2}.

The second technical result used without a proof in the main part of the paper was subadditivity of operational distance, i.e.,
\begin{equation}
D_{op}\rbracket{\M^{(1)}\otimes \M^{(2)},\N^{(1)}\otimes \N^{(2)}} \leq D_{op}\rbracket{\M^{(1)},\N^{(1)}} +D_{op}\rbracket{\M^{(2)},\N^{(2)}}\ ,
\end{equation}
where  $\M^{(1)}\otimes\M^{(2)}$ and $\N^{(1)}\otimes\N^{(2)}$ are POVMs on a compound quantum system $\mathcal{H}=\mathcal{H}_1 \otimes \mathcal{H}_2$ whose effects are build from single party measurements according to the following prescription:
\begin{equation}
\left(\M^{(1)}\otimes\M^{(2)}\right)_{(i,j)}\coloneqq M^{(1)}_i \otimes M^{(2)}_j \ .
\end{equation}
In order to prove \eqref{eq:subadditivity} we will again use 
Eq.~\eqref{eq::DopDtv} which yields
\begin{equation}\label{eq:int2}
D_{op}\rbracket{\M^{(1)}\otimes \M^{(2)},\N^{(1)}\otimes \N^{(2)}}= \max_{\rho} \frac{1}{2} \sum_{i,j}\left|\mathrm{tr}\left(\rho M^{(1)}_i \otimes M^{(2)}_j    - \rho N^{(1)}_i \otimes M^{(2)}_j \right)  \right| \ .
\end{equation}
Using triangle inequality we obtain that for all quantum states state $\rho$
\begin{equation}\label{eq:interm3}
\frac{1}{2} \sum_{i,j}\left|\mathrm{tr}\left(\rho M^{(1)}_i \otimes M^{(2)}_j    - \rho N^{(1)}_i \otimes M^{(2)}_j \right)  \right| \leq
\frac{1}{2} \sum_{i,j}\left|\mathrm{tr}\left(\rho M^{(1)}_i \otimes \left[ M^{(2)}_j - N^{(2)}_j\right]\right) \right|   + 
\frac{1}{2} \sum_{i,j}\left|\mathrm{tr}\left(\rho
\left[M^{(1)}_i- N^{(1)}_i\right] \otimes N^{(2)}_j \right)  \right|\ .
\end{equation}
To proceed we need to introduce auxiliary probability distributions and quantum states
\begin{equation}
p^{(2)}_i =  \mathrm{tr}\left( M^{(1)}_i \otimes \mathbb{I} \rho   \right)\ ,\ p^{(1)}_j =  \mathrm{tr}\left( \mathbb{I} \otimes N^{(2)}_j   \rho   \right)\ ,
\end{equation}
\begin{equation}
\rho^{(2)}_i = \mathrm{tr}_1 \left( \sqrt{M^{(1)}_i} \otimes \mathbb{I} \rho  \sqrt{M^{(1)}_i}\otimes \mathbb{I} \right)/p^{(2)}_i\ ,\ \rho^{(1)}_j = \mathrm{tr}_2 \left(  \mathbb{I} \otimes \sqrt{N^{(2)}_j}  \rho  \mathbb{I} \otimes \sqrt{N^{(2)}_j} \right) /p^{(1)}_j\ .
\end{equation}
After simple computations using these auxiliary objects we obtain 
\begin{equation}
\frac{1}{2} \sum_{i,j}\left|\mathrm{tr}\left(\rho M^{(1)}_i \otimes M^{(2)}_j    - \rho N^{(1)}_i \otimes M^{(2)}_j \right)  \right| \leq
\frac{1}{2} \sum_{i,j} p^{(2)}_i \left|\mathrm{tr}\left(\rho^{(2)}_i \left[ M^{(2)}_j - N^{(2)}_j\right]\right) \right|   + 
\frac{1}{2} \sum_{i,j} p^{(1)}_j\left|\mathrm{tr}\left(\rho^{(1)}_j
\left[M^{(1)}_i- N^{(1)}_i\right] \right)  \right|\ .
\end{equation}
Finally, directly form the definition of operational distance (see Eq.  \eqref{eq::DopDtv}) we obtain
\begin{equation}
\frac{1}{2} \sum_{i,j} p^{(2)}_i \left|\mathrm{tr}\left(\rho^{(2)}_i \left[ M^{(2)}_j - N^{(2)}_j\right]\right) \right| = \frac{1}{2} \sum_{i} p^{(2)}_i \sum_j \left|\mathrm{tr}\left(\rho^{(2)}_i \left[ M^{(2)}_j - N^{(2)}_j\right]\right) \right| \leq D_{op}\left(\M^{(2)},\N^{(2)}\right)
\end{equation}
and analogously 
\begin{equation}
\frac{1}{2} \sum_{i,j} p^{(1)}_j\left|\mathrm{tr}\left(\rho^{(1)}_j
\left[M^{(1)}_i- N^{(1)}_i\right] \right)  \right| \leq D_{op}\left(\M^{(1)},\N^{(1)}\right)\ .
\end{equation}
We conclude the proof by inserting the above two inequalities into the right hand side of Eq.~\eqref{eq:interm3}.

\section{State dependence of correction}\label{app::to_correct_or_not_to_correct}
In Section~\ref{sec: Main}, we developed a scheme for the mitigation of readout errors which relies solely on classical post-processing of the vector of experimental statistics.
We have shown that this procedure mitigates noise perfectly, provided the measuring device is affected by a certain type of classical noise.
In Section~\ref{sec: Errors}, we analyzed what effects on the mitigation scheme arises from the deviations from the adopted noise model and from the finite-size statistics.
In particular, we have pointed out that the procedure is highly on the quantum state being measured. Specifically, it might happen that correction actually worsens the results compared to the non-post-processed case.
To address this issue, we proposed a  criterion for deciding whether the mitigation has been successful (Eq.~\eqref{eq:RULEcorr}), which we will now rewrite here for convenience of the Reader
\begin{align}\label{eq::rule_correction_appendix}
\delta+\alpha < D_{op}\rbracket{\M^{\text{exp}},\M^{\text{ideal}}} + \epsilon \Rightarrow \text{mitigation succesful}\ .
\end{align}
In what follows we provide provide numerical arguments for approximate correctness of the above rule for single-qubit detectors and two-qubit uncorrelated detectors.

\begin{figure}[h!]
	\centering
	\includegraphics[scale=0.55]{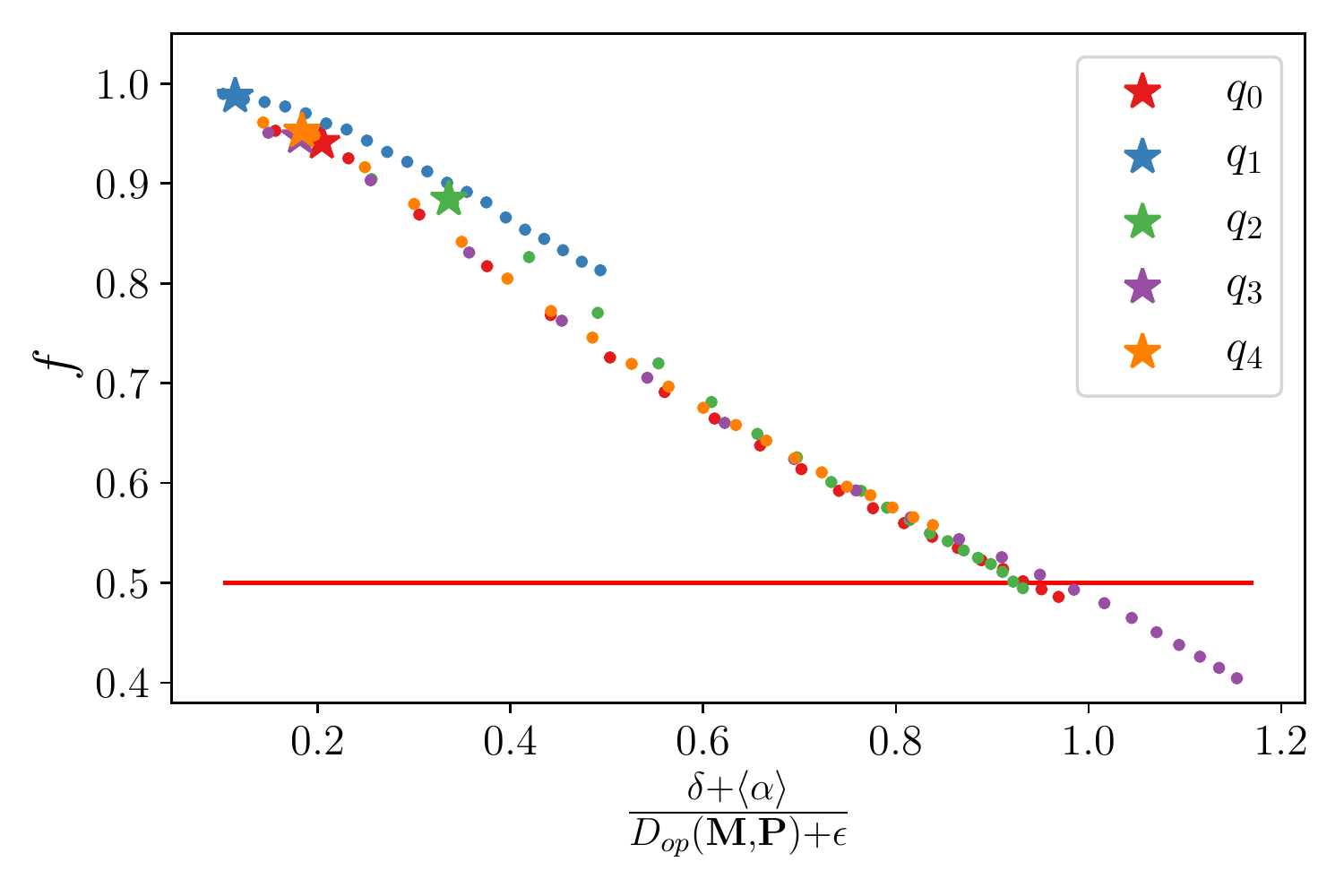}
	\caption{ \label{fig::1q_correct_or_not}
		Dependence between the fraction $f$ for which the mitigation is successful and the ratio $\frac{\delta+\left<\alpha\right>}{D_{op}\rbracket{\M,\P}+\epsilon}$.
		Each point corresponds to different non-classical part $z$ of the noise.
		For each point, fraction $f$ was calculated for $L=10000$ Haar-random quantum states. 
		The sampling size for the probability estimation was fixed at $N=8192$.
		The probability of error in the statistical part of the bound was fixed at $\mathrm{Pr}_{err}=0.01$. 
		The dependence of $\alpha$ on the particular case has been overcome by taking the average $\left<\alpha\right>$ over all ($L=10000$) implementations.
		Stars denote actual POVMs obtained via QDT on IBM's quantum device.}
\end{figure}

First, let us note that the above inequality is in fact comparison of two bounds.
On the LHS, we have the upper bound for the error for our mitigation procedure, i.e., the upper bound on TV-distance (Eq.~\eqref{eq::tv_distance}) between post-processed statistics $\vec{p}^{_\ast}_{\mathrm{exp}}$ and the ideal statistics $\vec{p}_{\mathrm{ideal}}$ that one would have obtained on the non-noisy detector.
On the RHS, we have an upper bound for TV-distance between non-post-processed statistics $\vec{p}^{\mathrm{est}}_{\mathrm{exp}}$, and the $\vec{p}_{\mathrm{ideal}}$. This is the \emph{worst possible error} one would expect on the noisy detector without mitigation.
Of course, in the real experiment, we do not have access to any particular $\vec{p}_{\mathrm{ideal}}$, which makes such general figures of merit particularly useful.
However, in the case of numerical simulations, when $\vec{p}_{\mathrm{ideal}}$ can be computed, we can use a quantum state sensitive criterion analogous to Eq.~\eqref{eq::rule_correction_appendix}, namely
\begin{align}\label{eq::rule_better_correction_appendix}
D_{TV}\rbracket{\vec{p}^{_\ast}_{\mathrm{exp}},\vec{p}_{\mathrm{ideal}}} < D_{TV}\rbracket{\vec{p}^{\mathrm{est}}_{\mathrm{exp}},\vec{p}_{\mathrm{ideal}}} \Rightarrow \text{mitigation succesful}\ .
\end{align}
It is now natural to ask how often the above criterion is satisfied, when Haar-random quantum states are measured by the noisy detector $\M$.
To address this question, we perform the following numerical procedure.
%Having the above rule in mind, we propose the following numerical procedure to estimate the chances of post-processing being beneficial for generic quantum states measured on the noisy detector $\M$ (which is assumed to be known, e.q., from detector tomography).
\begin{enumerate}	
	\item[1.] Generate $L$ Haar-random pure quantum states. \\
	For each quantum state:
	\begin{enumerate}
		\item[i.] Calculate probability vectors $\vec{p}_{\mathrm{exp}}$ and $\vec{p}_{\mathrm{ideal}}$, that quantum state generates, when measured by noisy detector $\M$ and ideal detector $\P$, respectively.
		\item[ii.] Sample $N$ times from the probability distribution $\vec{p}_{\mathrm{exp}}$, obtaining $\vec{p}^{\mathrm{est}}_{\mathrm{exp}}$.
		\item[iii.] Compute $\Lambda^{-1}\vec{p}^{\mathrm{est}}_{\mathrm{exp}}$  
		and check if it is physical probability vector. 
		If no, solve problem defined in Eq.~\eqref{eq::est_problem}, obtaining $\vec{p}^{_\ast}_{\mathrm{exp}}$ and calculate $\alpha$ defined in Eq.~\eqref{eq::error_strange}.
		If yes, identify $\vec{p}^{_\ast}_{\mathrm{exp}}=\vec{p}^{\mathrm{est}}_{\mathrm{exp}}$ and set $\alpha=0$.
		\item[iv.] Calculate $D_{TV}\rbracket{\vec{p}^{_\ast}_{\mathrm{exp}},\vec{p}_{\mathrm{ideal}}}$ and $D_{TV}\rbracket{\vec{p}^{\mathrm{est}}_{\mathrm{exp}},\vec{p}_{\mathrm{ideal}}}$. 
		If inequality in Eq.~\eqref{eq::rule_correction_appendix} is satisfied, the mitigation was successful in that case.
	\end{enumerate}
	\item[2.] Calculate fraction $f$ of quantum states for which mitigation was successful. 
	%	\item[3.] If $f>\frac{1}{2}$, then in general it is beneficial to use our mitigation scheme.
\end{enumerate}
We have studied how fraction $f$ changes with growing ratio $\frac{\delta+\alpha}{D_{op}\rbracket{\M,\P}+\epsilon}$ in the case of single-qubit detectors.
To this aim, we have implemented the above algorithm for the range of magnitudes of off-diagonal terms of POVM's elements, while keeping diagonal terms from actual IBM's data.
Results are shown in Fig.~\ref{fig::1q_correct_or_not}.
For all qubits the fraction of corrected statistics crosses $50\%$ in the region in which $\delta+\alpha \approx D_{op}\rbracket{\M,\P}+\epsilon$, which backs up rule give in Eq.~\eqref{eq::rule_correction_appendix}.
Furthermore, the fraction of successful error mitigations for actual experimental data lies between $88\%$ and $99\%$, which suggests that for single-qubit experiments our protocol should be helpful in most cases.

\begin{figure}[h!]
	\includegraphics[scale=0.245]{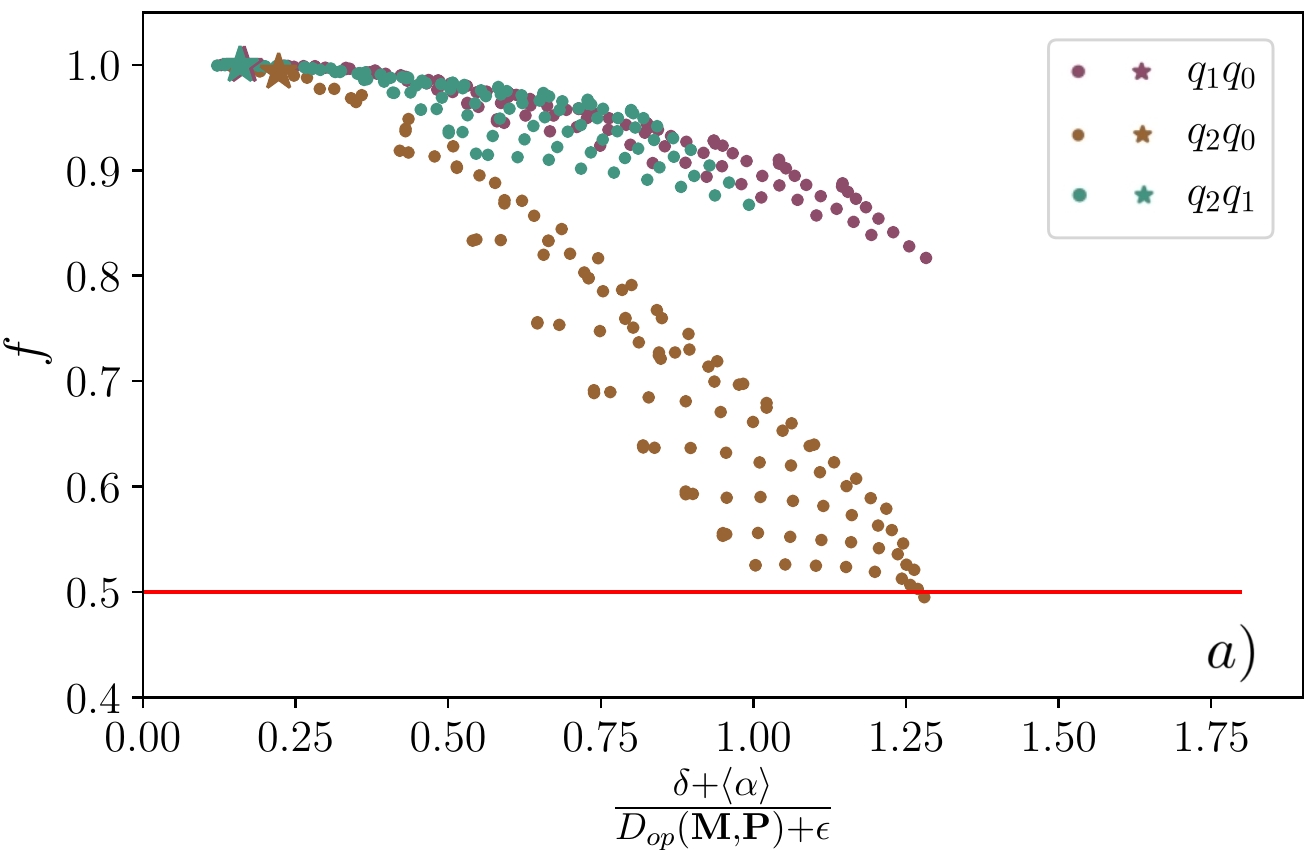}
	\label{fig::2q_fractions_a}
	\includegraphics[scale=0.248]{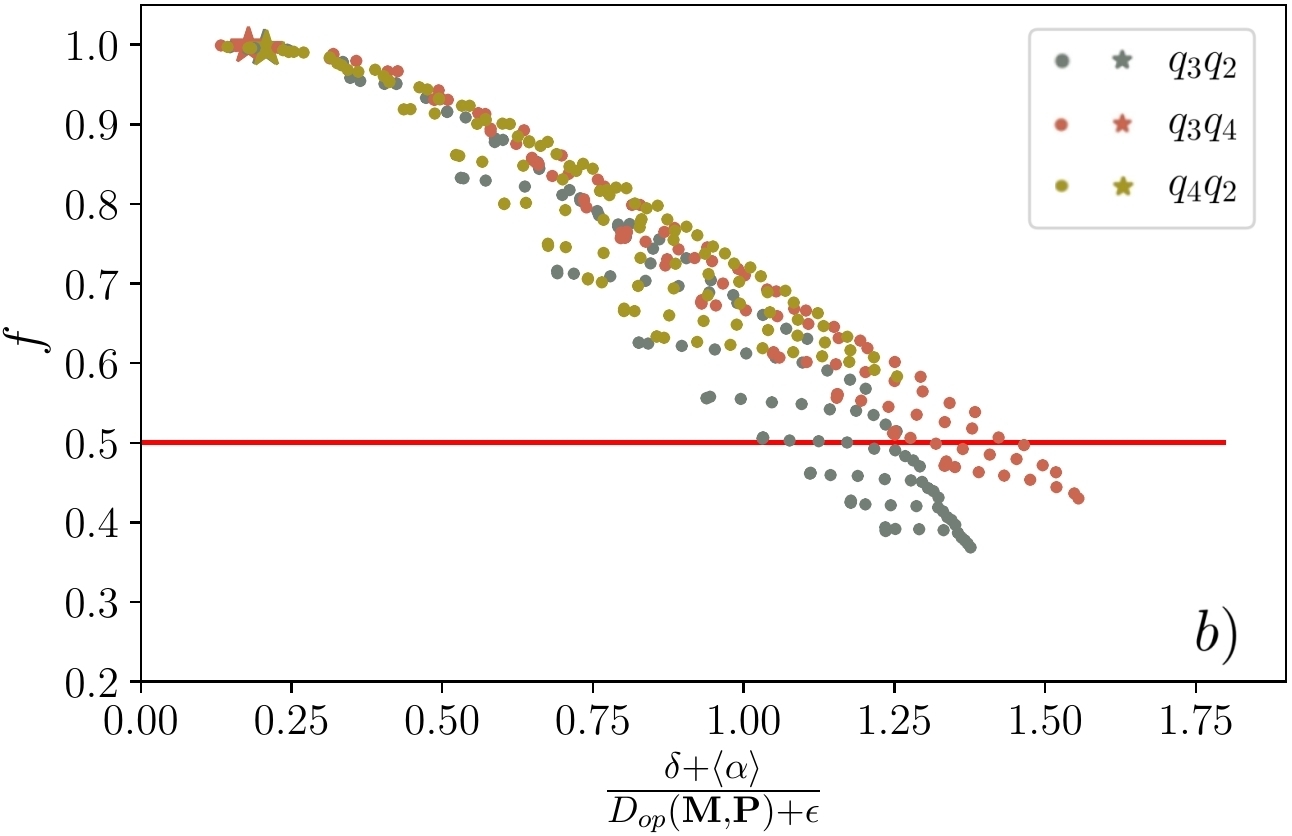}
	\label{fig::2q_fractions_b}\\
	\caption{ \label{fig::2q_fractions}	Dependence between the fraction $f$ for which the mitigation is successful and the ratio $\frac{\delta+\left<\alpha\right>}{D_{op}\rbracket{\M,\P}+\epsilon}$ .
		The data convention is fully analogous to that of Fig~\ref{fig::1q_correct_or_not}.
		Fraction $f$ was calculated for $L=10000$ Haar-random quantum states, each probability vector was sampled $N=8192$ times and the probability of error in the statistical part of the bound was fixed at $\mathrm{Pr}_{err}=0.01$.
		a) and b) correspond to different pairs of qubits, which were separated for clarity. POVMs were created by the tensor product of single-qubit measurements with increasing non-classical parts. 
	}
\end{figure}

For the two-qubit case we have tested POVMs created via the tensor product of single-qubit POVMs with increasing off-diagonal terms, again leaving diagonal terms from experimental data.
Results of simulation are shown on Figure~\ref{fig::2q_fractions}. 
Similarly to single-qubit case, fraction of successful mitigations crosses $50\%$ around the regime where Eq.~\eqref{eq::rule_correction_appendix} becomes unsatisfied.
Furthermore, for actual experimental data $f$ takes values between $98.86\%$ and $99.99\%$, which suggests that for \emph{uncorrelated} pairs of qubits our mitigation procedure should be successful for generic quantum states.

\section{Additional experimental data}\label{app::exp_data}
In the following section, we present explicit forms of matrices representing POVMs reconstructed in QDTs discussed in Section~\ref{sec: Validation}.
Furthermore, we present the dates of execution of all our experiments presented throughout the main part of the paper.
For further experimental data, we refer the Reader to the online repository \cite{online_repo}.

\subsection{Quantum Detector Tomography}
Here we present the explicit form of the first effects of the exemplary POVMs reconstructed in single-qubit detector tomographies on all five qubits of IBM's \textit{ibmqx4} and first five qubits of Rigetti's \textit{Aspen-4-16Q-A}. The second effect of each the POVMs can then be automatically obtained as the complement to identity.
\begin{center}
	a. IBM
\end{center}
\begin{align*}%
&M_1^{\mathrm{q_0}}= \begin{pmatrix}%
0.963  & 0.004  \\%
0.004  & 0.137  %
\end{pmatrix}\quad \ \ ,\ 
M_1^{\mathrm{q_1}}= \begin{pmatrix}%
0.99  & 0.002-0.001i \\%
0.002+0.001i & 0.37  %
\end{pmatrix},\\
&M_1^{\mathrm{q_2}}= \begin{pmatrix}%
0.986  & -0.001  \\%
-0.001  & 0.065  %
\end{pmatrix},\ 
M_1^{\mathrm{q_3}}= \begin{pmatrix}%
0.919  & 0.003-0.003i \\%
0.003+0.003i & 0.148  %
\end{pmatrix},\ 
\\
&M_1^{\mathrm{q_4}}= \begin{pmatrix}%
0.98  & -0.002i \\%
0.002i& 0.155  %
\end{pmatrix}.
\numberthis
\end{align*}%
\begin{center}
	b. Rigetti
\end{center}
\begin{align*}%
&M_1^{\mathrm{q_0}}= \begin{pmatrix}%
0.975  & -0.002  \\%
-0.002  & 0.124  %
\end{pmatrix}\qquad \qquad \qquad  ,\ 
M_1^{\mathrm{q_1}}= \begin{pmatrix}%
0.966  & 0.002+0.002i \\%
0.002-0.002i & 0.101  %
\end{pmatrix},\\
&M_1^{\mathrm{q_2}}= \begin{pmatrix}%
0.987  & 0.001-0.001i \\%
0.001+0.001i & 0.066  %
\end{pmatrix},\ 
M_1^{\mathrm{q_3}}= \begin{pmatrix}%
0.938  & 0.002+0.001i \\%
0.002-0.001i & 0.184  %
\end{pmatrix},\ 
\\
&M_1^{\mathrm{q_4}}= \begin{pmatrix}%
0.903  & 0.012-0.001i \\%
0.012+0.001i & 0.155  %
\end{pmatrix}.
\numberthis
\end{align*}%

\subsection{Dates of the experiments}

Table~\ref{tab::dates_ibm} contains the dates of execution of experiments performed on IBM's device \textit{ibmqx4}.
Fidelities on Rigetti's device given in Tab.~\ref{RB_Fig.} were checked via Forest \cite{ref_rigetti} on June 30, 2019, and all tomographic reconstructions on Rigetti's device were done on May 30, 2019.

\begin{table}[h!]
	\centering
	\begin{tabular}{@{}clc@{}}		
		\toprule			
		\multicolumn{1}{l}{}                                                           & Location \qquad     & Date\\	     	 \midrule
		%		\multirow{6}{*}
		& Tab.~\ref{RB_Fig.}  & February 15\\
		& Tab.~\ref{tab::correlations}  & April 28 \\
		& Tab.~\ref{tab::algorithms}  & March 12\\
		& Tab.~\ref{tab::PD_5q}  & May 30 \\
		& Fig.~\ref{fig::QDT_1q}  & April 28 \\
		& Fig.~\ref{fig::1q_arrows}  & April 28 \\
		& Fig.~\ref{fig::QDT_2q}  & April 28	 \\
		& Fig.~\ref{fig::QST_QPT_1q}  & April 28 \\
		& Fig.~\ref{fig::QST_2q}  & April 28 \\
		& Fig.~\ref{fig::naimark}  & February 21 \\
		\midrule
	\end{tabular}
	\caption{\label{tab::dates_ibm} The dates of execution of all our experiments performed on IBM's \textit{ibmqx4}. 
		We note that in the case of Tab.~\eqref{RB_Fig.} the data was obtained via qiskit \cite{qiskit_ref}.
		All experiments were done in 2019.}
\end{table}

\end{document}